\documentclass[sigconf]{acmart}
\usepackage{wrapfig}
\usepackage{enumitem}

\newcommand{\R}{\mathbb{R}}

\newcommand{\minisec}[1]{\noindent\textbf{#1.}}

\usepackage{xcolor}
\usepackage[linesnumbered,ruled,vlined]{algorithm2e}

\SetCommentSty{mycommfont}

\usepackage{multirow}

\usepackage{amsmath,amsfonts}
\usepackage{amsthm}

\usepackage{algorithmic}
\usepackage{textcomp}
\usepackage{balance} 
\usepackage{graphicx}
\usepackage{color}
\usepackage{soul}
\usepackage{lipsum}

\theoremstyle{definition}
\newtheorem{definition}{Definition}

\usepackage{subcaption}
\usepackage{graphicx}

\usepackage{xcolor}
\usepackage[normalem]{ulem}

\usepackage[T1]{fontenc}
\usepackage{listings}
\definecolor{codegreen}{rgb}{0,0.6,0}
\definecolor{codegray}{rgb}{0.5,0.5,0.5}
\definecolor{codepurple}{rgb}{0.58,0,0.82}
\definecolor{backcolour}{rgb}{0.95,0.95,0.92}
\definecolor{cadmiumgreen}{rgb}{0.0, 0.42, 0.24}
\AtBeginDocument{%
  \providecommand\BibTeX{{%
    \normalfont B\kern-0.5em{\scshape i\kern-0.25em b}\kern-0.8em\TeX}}}

\setcopyright{acmcopyright}
\copyrightyear{2018}
\acmYear{2018}
\acmDOI{10.1145/1122445.1122456}

\acmConference[Woodstock '18]{Woodstock '18: ACM Symposium on Neural
  Gaze Detection}{June 03--05, 2021}{Woodstock, NY}
\acmBooktitle{Woodstock '18: ACM Symposium on Neural Gaze Detection,
  June 03--05, 2018, Woodstock, NY}
\acmPrice{15.00}
\acmISBN{978-1-4503-XXXX-X/18/06}

\begin{document}

\title{Dolphin: An Actor-Oriented Database for Reactive Moving Object Data Management}



\author{Yiwen Wang}
\authornote{Work are done while author was affiliated with University of Copenhagen}
\email{wangyiwen0304@gmail.com}
\orcid{0000-0002-4531-2222}
\affiliation{
  \institution{Independent Researcher}
  \city{Copenhagen}
  \country{Denmark}
}

\author{Vivek Shah}
\authornote{Work partially done while author was affiliated with University of Copenhagen}
\authornotemark[1]
\email{bonii.vivek@gmail.com}
\affiliation{%
  \institution{Independent Researcher}
  \city{Copenhagen}
  \country{Denmark}
}

\author{Marcos~Antonio Vaz~Salles}
\authornote{Work are done while author was affiliated with University of Copenhagen}
\email{vmarcos@di.ku.dk}
\affiliation{%
  \institution{Independent Researcher}
  \city{Copenhagen}
  \country{Denmark}
}

\author{Claudia Bauzer Medeiros}
\email{cmbm@ic.unicamp.br}
\affiliation{%
    \institution{University of Campinas}
    \city{Campinas, SP}
    \country{Brazil}
}

\author{Julio Cesar Dos Reis}
\email{jreis@ic.unicamp.br}
\affiliation{%
    \institution{University of Campinas}
    \city{Campinas, SP}
    \country{Brazil}
}

\author{Yongluan Zhou}
\email{zhou@di.ku.dk}
\affiliation{%
  \institution{University of Copenhagen}
  \city{Copenhagen}
  \country{Denmark}
}

\begin{abstract}
Novel reactive moving object applications require solutions to support object reactive behaviors as a way to query and update dynamic data. 
While moving object scenarios have long been researched in the context of spatio-temporal data management, reactive behavior is usually left to complex end-user implementations.
However, it is not just a matter of hardwiring reactive constraints: the required solutions need to satisfy tight low-latency computation requirements and be scalable.
This paper explores a novel approach to enrich a distributed actor-based framework with reactive functionality and complex spatial data management along with concurrency semantics. 
Our approach relies on a proposal of the moving actor abstraction, which is a conceptual enhancement of the actor model with reactive sensing, movement, and spatial querying capabilities. This enhancement helps developers of reactive moving object applications avoid the significant burden of implementing application-level schemes to balance 
performance and consistency. Based on moving actors, we define a reactive moving object data management platform, named \textit{Moving Actor-Oriented Databases (M-AODBs)}, and build \textit{Dolphin} -- an implementation of M-AODBs.
Dolphin embodies a non-intrusive actor-based design layered on top of the Microsoft Orleans distributed virtual actor framework.
In a set of experimental evaluations with realistic reactive moving object scenarios, 
Dolphin exhibits scalability on multi-machines and provides near-real-time reaction latency.
\end{abstract}




\maketitle

\section{Introduction}
\label{sec:introduction}

In recent years, we are witnessing a growth in research in data-intensive Internet-of-Moving-Things scenarios involving mobile devices and sensors~\citep{DBLP:journals/tasm/Costa18,DBLP:journals/ijgi/CaoW20}. 
For instance, applications such as collaborative transportation systems~\citep{khajenasiri2017review, DBLP:journals/access/CuiSLZZ19}, fleet management and traffic monitoring~\citep{DBLP:journals/corr/abs-1806-06762,DBLP:conf/avss/AlbaniIHT17,smartfarmingIoT,DBLP:conf/icra/ShorinwaYHKS20, DBLP:conf/sii/HonkoteGNGS20,DBLP:conf/evoW/AttenCDB16}, tracking of mobile objects and crowd sensing platforms \citep{DBLP:journals/comsur/CapponiFKFKB19}, \textit{etc.}, concern objects that continuously move in space and react online to their surroundings. 
 
Reactivity is a crucial new requirement for these applications, \textit{e.g.}, as objects
change their states and share information 
to affect other ``surrounding'' moving objects. We name such objects \emph{reactive moving objects} and these applications \emph{reactive moving object applications} -- to characterize the emphasis on surroundings-sensitive object behavior, and the collective participation of such objects in a spatially-sensitive decision-making process.

\setlength{\columnsep}{-0cm}%
\begin{wrapfigure}{l}{0.22\textwidth}
 \vspace{-0.3cm}
  \includegraphics[scale=0.105]{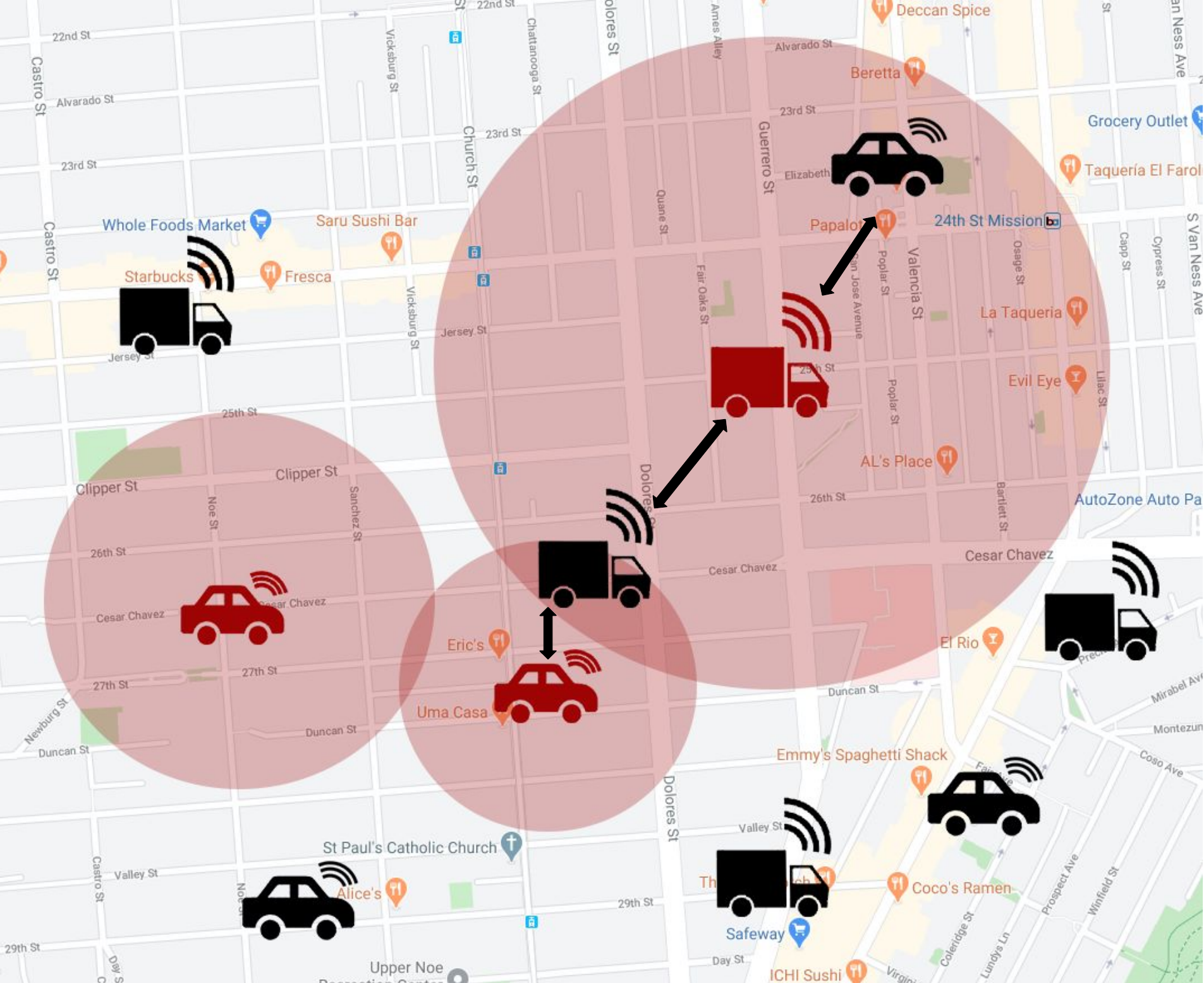}
  \vspace{-0.3cm}
  \caption{C-ITS Example}
  \label{fig:citsexample}
  \vspace{-0.3cm}
\end{wrapfigure}

\minisec{Running Example} In Cooperative Intelligent Transportation Systems (C-ITS) \citep{DBLP:journals/corr/abs-2011-03425, bussche2020hedging, DBLP:conf/mdm/NguyenHSE20}, vehicles communicate and cooperate towards improving overall transportation effectiveness, driving behaviours, security and safety \citep{ni2016traffic}. While regular ITS~\citep{DBLP:journals/itsm/LiN13, DBLP:journals/ral/HuBALL20, tak2020sectional} concentrate on local information sharing, optimization, and coordination (\textit{e.g.}, involving drivers, traffic controllers, or transportation system operators) to make better decisions, in C-ITS, global information ingestion is required and vehicles themselves are active participants in transportation actions and decisions. Figure~\ref{fig:citsexample} presents an example of C-ITS. Reactive vehicles (shown as red vehicle icons) start reactive sensing within a specific spatial region (shown as a semi-transparent red range). A reactive vehicle is aware of other vehicles if those vehicles' movements satisfy its predefined spatial predicates (\textit{e.g}., cross, cover, or overlap) against its spatial sensing area. Sensing areas and predicates may vary from vehicle to vehicle. 
In addition, reactive vehicles can take reactive actions, such as communication with approaching vehicles to send warnings to help them avoid hazards~\citep{Uhlemann18:ConnectedVehicles}. We envision cooperation will be a critical ingredient for self-driving or assisted-driving.$\qed$ 

According to~\citet{DBLP:journals/taosd/SalvaneschiM13} as well as~\citet{boner2014reactive}, a reactive system needs to update and react to the changes of inputs correspondingly in tight real-time computation constraints. This is challenging to meet in data-intensive settings. In conventional multi-tier distributed data architectures, frequent data shipment is required between the stateless application tier running the application codes and the database tier storing the application state. In addition, system performance would be constrained by the network bandwidth and latency. One might suggest to push the whole logic of a reactive moving object application down into a spatial DBMS. However, reactive features for moving objects are lacking in these systems. Triggers are the primary reactive abstraction available, but they are riddled with complex semantic issues for developers~\citep{widom1996active}. Furthermore, there are concerns regarding trigger scalability~\citep{DBLP:conf/icde/HansonCHKNPPV99}, which are only made worse in highly dynamic and distributed environments.

Distributed in-memory application architectures have proved to be a viable solution to address the aforementioned performance challenges~\citep{DBLP:books/daglib/0066897, DBLP:journals/internet/BernsteinB16}. These architectures often leverage the actor model as a programming abstraction for middle tier implementation~\citep{hewitt2010actor,karmani2011actors}, while employing distributed DBMS for selective persistence of actor state~\citep{bernstein2014orleans}. Application logic and data are encapsulated into actors, each of which naturally represents the digital twin of a real-world object~\citep{DBLP:conf/edbt/WangRBSMZ19,DBLP:conf/icde/Bernstein18}. The decomposition of application logic and data into actors, as well as the distributed and concurrent actor model execution enable efficient use of computational resources to boost scalability.



\begin{sloppypar}
In light of the success of the actor model in building high-throughput and low-latency applications, we envision it as a compelling implementation abstraction for the application tier in reactive moving object applications -- each reactive object, as the digital twin of a real-world physical reactive moving entity, can be modeled as an actor. However, present actor runtimes do not have any built-in support for representing movement of objects in space, for querying their locations, or, most importantly, for specifying reactive behaviors to be triggered in response to the movement of other objects. Developers must face a choice between either: (a) scalably implementing these challenging reactive and spatial functionalities in the application tier; or (b) weaving together a complex web of systems, \textit{e.g.}, spatial streaming platforms, spatial DBMS, and distributed mid-tier frameworks, with little support to navigate the trade-offs in  cross-system data consistency semantics.
\end{sloppypar}

In this article, we investigate how to design and build an actor-based data platform with high-throughput, low-latency, and reactive spatial data management functionalities. 
Our solution supports the development of application tiers for data-intensive reactive moving object applications. 
Our platform alleviates the burden on application developers by allowing them to concentrate on their application logic instead of being distracted by complex spatial data management tasks. In line with the terminology in~\citep{DBLP:conf/icde/Bernstein18}, we call such a platform a Moving Actor-Oriented Database (M-AODB).

There are several major challenges to achieve this goal. 
First, it is necessary to provide a proper abstraction for representing reactive moving objects. In this sense, the classic actor model needs to be enriched with features required by reactive moving object applications, including geo-referenced attributes, spatial indexing and querying, spatial-driven reactive behaviors, etc. 
Second, in the actor model, actors update and query data in a concurrent and asynchronous manner. In our context, concurrency semantics have to be defined, not only to provide proper data consistency guarantees to developers, but also to achieve trade-offs between data latency and consistency. 
Third, these additional features need to be offered in a manner that preserves the benefits of scalability and low latency that actor runtimes exhibit for conventional, non-spatial applications.
A fourth major challenge is the coherent design and implementation of a M-AODB that facilitates distributed deployment of reactive moving object applications in the cloud. 

Our research addresses these challenges through Dolphin, a concrete implementation of an M-AODB. To the best of our knowledge, 
Dolphin is the first distributed in-memory spatial data platform that can be deployed on the cloud 
\emph{and} supports scalable and reactive spatial data management for reactive moving object applications. 

The major contributions of our article thus include the following:
\begin{itemize}[leftmargin=*]
    \item In Section~\ref{sec:m-aodb}, we propose a \textit{Moving Actor} abstraction, which brings together the actor model with reactive spatial data management. 
    Moreover, we propose a novel architecture named \textit{Moving Actor-Oriented Databases (M-AODBs)} that supports the proposed moving actor model. In this context, a M-AODB is a reactive, stateful, and scalable distributed actor-oriented data platform. 
    
    \item In Section~\ref{sec:semantics}, 
    we define and implement two concurrency semantics for the spatial data of reactive moving actors, namely \textit{Actor-Based Freshness semantics} and \textit{Actor-Based Snapshot semantics}. They are essential to provide concurrency semantics for developers to reason about correctness and performance, as well as to achieve trade-offs between data latency and consistency.

    \item  In Section~\ref{sec:dolphin}, we present \textit{Dolphin} -- our 
    cloud-ready implementation of the conceptual proposal of M-AODBs on top of the virtual actor framework Microsoft Orleans~\citep{bernstein2014orleans}. 
    In Dolphin, the functionalities of reactive spatial data management were themselves implemented using actors over space partitions to both benefit from the underlying framework and achieve scalability. 
    We present how these actors interact asynchronously to fulfill both the 
    Freshness and Snapshot concurrency semantics.
    
    \item In Section~\ref{sec:experimentalevaluation}, we experimentally evaluate Dolphin on synthetic and realistic datasets to showcase the capabilities of our design and implementation of M-AODBs. Our results show that Dolphin supports low-latency reactions, frequent data updates and queries, and scales over multiple machines. 
\end{itemize}

Related work is discussed in Section \ref{sec:relatedwork}, and we conclude this paper in Section~\ref{sec:conclusion}.

\vspace{-0.2cm}
\section{Towards Moving Actor-Oriented Databases}
\label{sec:m-aodb}

\subsection{Design Objectives}
\label{sec:reqs}
Our goal is to develop a data platform for application tier development for the aforementioned novel reactive moving object applications. We summarize the high-level design objectives as follows:

    \noindent\textbf{O1. Support Non-reactive Spatial Data Management}. Conventional spatial data management functionalities like data updates, spatial indices and queries, and maintenance of spatial static integrity constraints are necessary to build these applications.

    \noindent\textbf{O2. Support Spatial Reactive Behavior.} Reactive objects must be able to sense their surroundings or other areas of space to perform reactive actions. This requires efficient processing of reactive actions of every object triggered by its surrounding peer objects in the environment. 
    
    \noindent\textbf{O3. Support Concurrent Heterogeneous Moving Objects.} To achieve low latency, the updates and reactive operations of moving objects should be executed concurrently. The data platform should provide a concurrent programming abstraction. Moreover, different moving objects could have customized and heterogeneous behaviors. So the abstraction should allow modeling objects by their types and behaviors. 
    
    \noindent\textbf{O4. Provide Scalability and Elasticity.} The platform should scale to arbitrary numbers of objects while supporting the processing of complex space-dependent application logic. Additionally, a reactive moving object is a digital twin of a real-world entity. Its lifecycle should be the same as that of the corresponding real world entity as it enters and/or leaves the system. Since load may vary dynamically depending on the relative positions of moving objects, the system should have the possibility to manage 
    resources elastically. 
    
\subsection{The Moving Actor Abstraction} 
\label{sec:moving:actor}

To achieve the above objectives, we start by looking at the programming abstraction of our data platform.
Among the programming abstractions of reactive applications, the actor model represents a distinctive choice \citep{hewitt2010actor, karmani2011actors, gupta2012akka}. 
In particular, the \emph{virtual actor} abstraction introduced by~\citet{DBLP:conf/cloud/BykovGKLPT11}, considers actors as modular and stateful virtual entities in perpetual existence, which facilitates a one-to-one mapping to moving objects in our context. Since virtual actors always exist, they do not require management of actor lifecycle and failures, which enhances developer productivity.
Virtual actors are a natural fit  to meet the scalability, availability, and elasticity challenges of emerging reactive moving object workloads, since they can be automatically and dynamically activated or deactivated and replicated to handle demand while balancing load across servers~\citep{DBLP:conf/icde/Bernstein18, bernstein2014orleans}.
Moreover, the concerns of fault tolerance and elasticity are managed by the runtime, which makes virtual actors a very intuitive and developer friendly programming abstraction. The widespread deployment of the virtual actor runtime \textit{Microsoft Orleans} in a variety of highly available and scalable low-latency production cloud services highlights the popularity of the programming model as well as the performance and maturity of the system~\citep{bernstein2014orleans}.

Therefore, we consider virtual actors to be a promising abstraction for managing reactive moving object data, as demonstrated by its successful and wide deployment in reactive applications. 
However, the actor model does not provide features that we require: (a) support for geo-referenced attributes and defining spatial and reactive functionalities in addition to user-defined methods; and (b) reacting to actions from other objects in space or information from the environment, 
\textit{e.g.}, updating its state or building connections with other moving objects.
To fill this gap, we propose the novel abstraction of \emph{moving actors}, which integrates the actor programming model with reactive moving object features.


%

\subsubsection{Moving Actor Formalization} \label{sec:movingactorform}
An \emph{actor} is a computational entity that keeps its private state and can only modify other actors' states by communicating via immutable asynchronous messages \citep{DBLP:books/daglib/0066897}. To extend the actor model with the two required features above, we formalize a moving actor as follows: 

\theoremstyle{definition}
\vspace{-1ex}
\begin{definition}
A \textbf{Moving Actor $a$ $(id, P, M)$} is an actor comprising of the following characteristics:
\begin{enumerate}[leftmargin=*]
    \item A unique identity $id$ to identify the moving actor $a$;
    
    \item Properties $P$ of $a$ containing spatial information, namely: (a) the current known location $l$ of this moving actor, where $l=(x,y) \in \R^{2}$,\footnote{$\R^{2}$ refers here to a two-dimensional real-valued space under a projected geospatial coordinate system, while $x, y$ are respectively latitude and longitude.} and (b) a fence $f$ around location $l$ representing a 
    polygon 
    to define its spatial sensing boundaries.
    
    \item Methods $M$ including:
    \begin{itemize}[leftmargin=-0.2cm]
         \item \texttt{Move($l_d$)}: Updates the location property of $a$ from its current location $l$ to a next location $l_d$. 
         The movement of $a$ generates a corresponding new fence $f$ that moves along with it. Besides, $a$'s movement may trigger reactive functions in other moving actors upon satisfaction of the spatial predicates associated to their fences (see below). 
       
        \item \texttt{FindActors($q$)}: Given a spatial query $q$, returns the actors that satisfy $q$. As a proof of concept, we focus on spatial range query.  
        Here, $q$ contains a spatial range $r$, where $r$ is a regular quadrilateral, s.t. $r$= $\left \{ \right. $ $(x_{min}, y_{min})$, $(x_{max}, y_{max})$ $\left. \right \}$ $\in \R^{2}$. This method can be extended with other types of spatial queries, such as KNN.
      
        \item \texttt{StartReactiveSensing($p$, $m$)}: Enables the sensing behavior of the moving actor. $a$ calls $StartReactiveSensing(p,m)$ to start sensing the environment using a spatial predicate $p$, e.g., cross, cover, overlap~\citep{clementini1995comparison}, which tests whether any moving actor's itinerary crosses, is covered by, overlaps the $f$ of $a$ respectively. If the movement of another moving actor satisfies $p$ with respect to $a$'s fence $f$, an application-defined reactive method $m$ in $a$ is invoked; $m$ encodes the reactive behaviors of $a$.
        
        \item \texttt{EndReactiveSensing()}: Disables the environment sensing behavior of the moving actor.
    \end{itemize}
\end{enumerate}
\label{def:moving:actor}
\end{definition}
\vspace{-2ex}

\subsubsection{Moving Actors in the C-ITS Example}
\label{subsubsec:citsexample}
In C-ITS, each vehicle $i$ can be modeled as a moving actor $a_i$. Accordingly, properties $a_i.P$ contain the current known location of the vehicle and its fence to enable it to sense and react to the movement of other nearby vehicles. $a_i.P$ can be further augmented with application-specific properties, such as the vehicle's weight or class.  

As vehicle $i$ moves in the physical world, $a_i$.\texttt{Move($l_d$)} is called to update its location and fence. This movement may trigger reactive functions of other moving actors. 
Moreover, while $i$ moves, it may have a series of complex behaviors, depending on the spatial events on its path. Consider the scenario where the vehicle finds a hazard (e.g., aquaplaning) as it goes along a road. Not only does it want to warn approaching vehicles, but also those that it will encounter later along its trajectory. This will require not only reactive behavior in the present moment (when it finds the hazard), but also in the future over a period of time. This can be achieved by taking advantage of the reactive sensing feature with reactive methods encoding the desired reactive behavior.

\begin{sloppypar}
If the reactive behavior of $i$ is to warn vehicles in the hazard's proximity, $a_i$ can invoke \texttt{FindActors($r$)} to find out which other vehicles are already in a spatial range $r$ around it and directly send a message to the corresponding moving actors about the hazard -- a single reaction to its present spatial context.
$a_i$.\texttt{StartReactiveSensing($p$,$m$)}, on the other hand, can be used for subsequent future (and continuous) behavior.
Here, $p$ is a cross spatial predicate to identify other vehicles that cross  $i$'s path as it continues moving forward.
Whenever this happens, $m$ will notify the corresponding digital twin about the hazard.
Finally, when $i$ is too far from the hazard for the warning to be effective, $a_i$.\texttt{EndReactiveSensing()} can be invoked.
\end{sloppypar}

\subsubsection{Moving Actors vs. Moving Objects}
It is noteworthy to point out the differences between the concepts of moving actor and moving objects as in spatiotemporal databases. Moving objects can be seen as data \citep{DBLP:journals/geoinformatica/Brinkhoff02}, over which operations are conceptually updates or queries from front-ends to back-ends on database representations of the moving objects. By contrast, moving actors compose data, operations, and reactive behaviors. They are not only used for moving object data registration and retrieval, but also to encode independently running digital twins of moving objects. As a key aspect, reactive moving objects modeled as moving actors interact with their surrounding objects and execute reactive functions themselves. For instance, in C-ITS, every individual vehicle can be modeled as a moving actor, which maintains its current location and can send queries to get information on other moving actors. Also, as exemplified in Section~\ref{subsubsec:citsexample}, moving actors can take reactive actions based on other moving actors' behaviors.

\subsection{Architecture of an M-AODB}
\label{sec:m-aodb:architecture}

\begin{wrapfigure}{l}{0.27\textwidth}
\vspace{-0.2cm}
 \includegraphics[scale=0.5]{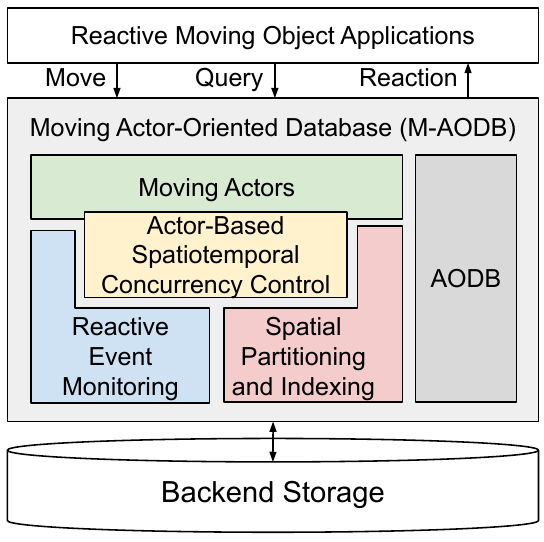}
  \vspace{-0.4cm}
  \caption{Architecture of an \\M-AODB}
  \label{fig:frameworkstructure}
  \vspace{-0.3cm}
\end{wrapfigure}

To achieve our design objectives, we build upon the concept of Actor-Oriented Databases (AODBs) \citep{DBLP:conf/icde/Bernstein18}, and propose \textit{Moving Actor-Oriented Database (M-AODB)}, a data platform for stateful, scalable, elastic, and distributed reactive moving object data management. Figure~\ref{fig:frameworkstructure} depicts the conceptual architecture of an M-AODB, highlighting its key components, which we describe below. 

\minisec{AODB}
To leverage all the advantages of an AODB, an M-AODB utilizes an AODB as a building block, and enriches the virtual actor abstraction to realize moving actors. 
We present one possible concrete realization of this integration in Section~\ref{sec:dolphin:design}, wherein the M-AODB's functionality is built into the virtual actors of AODBs. 

\minisec{Moving Actors}
Reactive moving object applications interact with M-AODBs by leveraging the abstraction of moving actors (Definition~\ref{def:moving:actor}, Section~\ref{sec:moving:actor}). Application-defined actors extend moving actors in our abstraction, enabling them to process moves, spatial range queries, and reactive function invocations, termed henceforth \emph{reactions}. For brevity, the data transfer stage from physical entities to their application-defined digital twins is not included in our architecture and further discussion.
 

\minisec{Actor-Based Spatiotemporal Concurrency Control}
Moves, spatial range queries, and reactions in M-AODBs are invoked concurrently by distinct distributed moving actors. Since the logic in each actor is conceptually isolated, actor-based spatiotemporal concurrency control semantics is needed to clearly specify the visibility of movement of other moving actors within each moving actor operation. Building on the moving objects literature, we develop two such actor-based semantics, described in Section~\ref{sec:semantics}. 

\minisec{Reactive Event Monitoring} 
A move operation by a moving actor may dynamically generate reactions on other moving actors that are currently observing the spatial region where the movement took place. 
In M-AODBs, this is achieved by reactive event generation and matching. Section~\ref{sec:move:reaction:workflows} gives an overview of the mechanism used in our implementation. 

\minisec{Spatial Partitioning and Indexing} 
The spatial data systems literature has shown repeatedly that spatial awareness is a pre-requisite for achieving scalability behavior over spatial workloads~\citep{DBLP:conf/icde/EldawyM15}. In M-AODBs, spatial partitioning can be leveraged to drive inter- and intra-machine parallelism in move and reaction processing as well as -- combined with spatial indexing -- to support spatial range queries (as we verify for our implementation in Section~\ref{sec:experimentalevaluation}).

Using the components described above, M-AODBs inherit the benefits of AODBs and extend it with support for spatial data management and reactivity.

\section{Actor-Based Spatiotemporal Concurrency Semantics of M-AODBs}
\label{sec:semantics}

In an M-AODB, moves, queries, and reactions take place concurrently in distinct distributed moving actors. Even though moving actors can be seen as isolated processes, their operations affect each other due to data dependencies. Therefore, a semantics needs to be defined to guarantee the correct result of concurrent operations on moving actors. 
M-AODBs need to decide at which identified points of time to make actor updates visible. 
Based on the analysis of reactive moving object application use scenarios, we specify and implement two concurrency semantics, \emph{Actor-Based Freshness Semantics} and \emph{Actor-Based Snapshot Semantics}, which are offered by M-AODBs and recommended on application level. 

\subsection{Actor-Based Freshness Semantics}
The Freshness semantics, originally proposed by~\citet{vsidlauskas2012parallel}, always provides fresh results of moving objects. Due to the asynchronous messaging in M-AODBs, we adapt it to provide ``fresh" (i.e., with a recency guarantee) results of moving actors. More precisely, we define the Actor-Based Freshness Semantics as:

\begin{definition}[Adapted from~\citep{vsidlauskas2012parallel}]
\label{def:findactor_fresh}
For a range query $FindActors(r)$ concurrently executed with movement $Move(l_d)$, assume the processing of the query lasts from $t_s$ to $t_e$, the movement update completion time of a moving actor $a$ in an M-AODB is $t_u$, and $a$ moves from $l_s$ to $l_d$. Assume further for any $a$ that it can only be updated at most once during $[t_s, t_e]$.\footnote{This assumption is made in~\citep{vsidlauskas2012parallel} to avoid keeping a history of move locations similarly to MVCC.} 
The result $A$ of $FindActors(r)$ satisfies the Actor-Based Freshness semantics if, for any moving actor $a$, the following holds:

\begin{itemize}
    \item if $a.t_u$ $<$ $t_s$ then $a$ $\in$ $A$ if and only if $a.l_d$ $\in$ $r$.
    \item if $t_s$ $<$ $a.t_u$ $<$ $t_e$ then:
    \begin{itemize}
        \item if $a.l_s$ $\in$ $r$ and $a.l_d$ $\in$ $r$, then $a$ $\in$ $A$. 
        \item if $a.l_s$ $\notin$ $r$ and $a.l_d$ $\notin$ $r$, then a $\notin$ $A$. 
        \item if $a.l_s$ $\in$ $r$ and $a.l_d$ $\notin$ $r$, then $a$ may or may not belong to $A$.
        \item if $a.l_s$ $\notin$ $r$ and $a.l_d$ $\in$ $r$, then $a$ may or may not belong to $A$.
    \end{itemize}
\end{itemize}
\end{definition}

Query results under Actor-Based Freshness semantics are dependent on the query processing time. That is because when queries are being processed, updates are also carried out concurrently, so the query results might or might not contain such concurrent updates depending on when the data are accessed by the query executor. Assume the processing of a query starts at $t_s$ and ends at $t_e$. With the Actor-Based Freshness semantics, query results would reflect all updates of moving actors that have been completed before $t_s$, and some of the fresher updates of moving actors that are completed between $t_s$ and $t_e$. 

\begin{wrapfigure}{l}{0.25\textwidth}
  \vspace{-0.1cm}
  \includegraphics[scale=0.6]{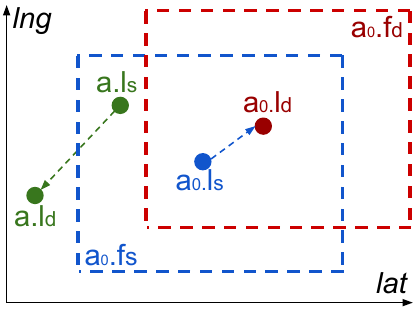}
  \vspace{-0.3cm}
  \caption{Actor-Based Freshness Semantics for StartReactiveSensing}
  \label{fig:freshnessstartreactivesensing}
  \vspace{-0.3cm}
\end{wrapfigure}

While there is no prior semantics defined for reactive actions in reactive object data management systems, we extended Actor-Based Freshness semantics to reactive behavior in M-AODBs. Figure~\ref{fig:freshnessstartreactivesensing} presents the situation that moving actors $a_0$ and $a$ are in a two-dimensional space $\R^{2}$, where $a_0$ is defined with a continuously updated fence $f$ under spatial predicate $p$ and a reactive method $m$. After $a_0$ issues $StartReactiveSensing(p,m)$, if any other moving actors satisfy spatial predicate $p$ against its fence, method $m$ of $a_0$ will be triggered. For example, $p$ can be set to detect if any moving actor $a$'s movement $cross$ $a_0$'s fence. If true, method $m$ is triggered to build a connection with $a$ and to share the information $a_0$ has obtained. We provide a formal definition for $StartReactiveSensing(p, m)$ under Actor-Based Freshness semantics as follows.

\begin{definition}
\label{def:sensing_fresh}
Given $a_0.StartReactiveSensing(p, m)$ is invoked, assume the latest move of $a_0$ is at $a_0.t_u$ from $a_0.l_s$ to $a_0.l_d$. 
The sensing fences of $a_0$ before and after this move are $a_0.f_s$ and $a_0.f_d$, respectively. Assume further that for any moving actor $a$, a move is at $a.t_u$ and the itinerary of this move is $a.iti$=\{$a.l_s$, $a.l_d$\}. Analogously to Definition~\ref{def:findactor_fresh}, we assume for any such $a_0$, it can only be updated at most once during the time needed to process a move of $a$ to eliminate checking predicate over multiple fences.\footnote{The given assumption is not guaranteed in realistic reactive moving object applications setting, but it is true for most real cases. In the C-ITS example, given a fast speed of moving object is 80 km/h and a required high accuracy of 5 meters, move time is around 225 milliseconds, which is longer than the processing time of move in an in-memory setting. Relaxing this assumption is an extension that we leave to future work.} A reactive action satisfies Actor-Based Freshness semantics if the following holds:
\begin{itemize}[leftmargin=*]
    \item When $a_0.t_u$<=$a.t_u$, the reactive method $m$ in $a_0$ is triggered if $a.iti$ satisfies spatial predicate $p$ against $a_0$.$f_d$,
    \item When $a_0.t_u$>$a.t_u$, 
        \begin{itemize}
            \item the reactive method $m$ in $a_0$ is triggered if $a.iti$ satisfies spatial predicate $p$ against $a_0$.$f_s$ and $a_0$.$f_d$.
            \item the reactive method $m$ in $a_0$ will not be triggered if $a.iti$ does not satisfy spatial predicate $p$ against neither $a_0$.$f_s$ nor $a_0$.$f_d$.
            \item the reactive method $m$ in $a_0$ may or may not be triggered if $a.iti$ satisfies spatial predicate $p$ against only $a_0$.$f_s$ or $a_0$.$f_d$.
        \end{itemize}
\end{itemize}
\end{definition}
  
\subsection{Actor-Based Snapshot Semantics}
As mentioned earlier, an M-AODB is an asynchronous distributed system. In order to satisfy applications with operations on a global state of an asynchronous distributed system, we define a Actor-Based Snapshot semantics in M-AODBs. Under this semantics, a global snapshot of the system at an approximate point in time is given. In contrast to the Actor-Based Freshness semantics, where the length of a query affects the time difference between locations returned, the results of operations depend only on the snapshot, thus making the processing time of operations irrelevant. The snapshot of the distributed application is updated periodically. 

We utilize loosely synchronized clocks to facilitate checkpoint coordination in M-AODBs. In single data center cloud distributed nodes, loosely synchronized clocks differ at most by an acceptable skew~\citep{DBLP:conf/sigmod/AdyaGLM95}, which can be enforced by the Network Time Protocol~\citep{DBLP:journals/rfc/rfc1305}. We argue that this is acceptable for reactive moving object applications, since minor differences in real time cannot be effectively observed in physical reality.

\begin{figure}[!ht]
  \centering
   \vspace{-0.1cm}
  \includegraphics[scale=0.9]{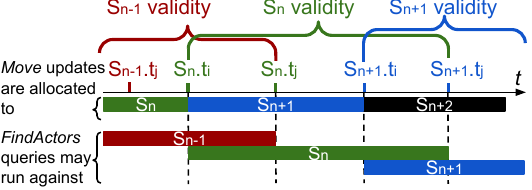}
   \vspace{-0.3cm}
  \caption{Actor-Based Snapshot Semantics: Queries read from active snapshot as of their start time. Updates are guaranteed to be reflected if they precede the start of the construction of the next snapshot.}
  \label{fig:snapshotmove}
  \vspace{-0.2cm}
\end{figure}

In M-AODBs, moving actors store update information locally. The loosely synchronized clock in each moving actor triggers its local action of exposing local state at approximately the same time to build a global snapshot in the distributed system~\citep{DBLP:conf/sigmod/AdyaGLM95, DBLP:conf/srds/DuEZ13}. Actor-Based Snapshot semantics, outlined in Figure \ref{fig:snapshotmove}, is defined as follows:

\begin{definition}
\label{def:findactors_snap}
Assume the time of starting and completing the construction of a snapshot $S_n$ is denoted by $S_n.t_i$ and $S_n.t_j$, respectively. $S_n$ should start being constructed after the construction of the last snapshot $S_{n-1}$ is finished, where $S_{n-1}.t_j$ $\leq$ $S_n.t_i$. $S_n$ completely replaces $S_{n-1}$ at the time $S_n.t_j$. For any moving actor $a$, assume $a$ moves to $l_d$ at time $t_u$. The snapshot $S_n$ satisfies the following:
    \begin{itemize}
        \item $a.l_d$ $\notin$ $S_n$ if $a.t_u$ $\geq$ $S_n.t_i$.
        \item $a.l_d$ $\in$ $S_n$ if $a.t_u$ $<$ $S_n.t_i$.
    \end{itemize}
    
Given a range query $FindActors(r)$ starts at time $t_s$, the result of $FindActors(r)$ satisfies the Actor-Based Snapshot semantics if the following holds:
\begin{itemize}
    \item if $t_s$ $\leq$ $S_n.t_i$, $FindActors(r)$ is executed over $S_{n-1}$.
    \item if $S_n.t_i$ $<$ $t_s$ $\leq$ $S_n.t_j$, $FindActors(r)$ is executed over either $S_{n-1}$ or $S_n$.
    \item if $S_n.t_j$ $<$ $t_s$, $FindActors(r)$ is executed over $S_n$.
\end{itemize}
\end{definition}

In addition to formalizing the Actor-Based Snapshot semantics as above, we also extend it to reactive actions in M-AODBs. Figure~\ref{fig:snapshotstartreactivesensing} illustrates two moving actors $a_0$ and $a$ in a two-dimensional space $\R^{2}$, where $a_0$ is doing reactive sensing with fence $f$, spatial predicate $p$ and reactive function $m$. An M-AODB periodically updates snapshots and a snapshot $S_n$ starts being generated at $S_n.t_i$. 
The accumulated spatial fence of $a_0$ is a minimum convex polygon covering all the fences of all the locations of $a_0$ between $S_{n-1}.t_i$ and $S_n.t_i$. In other words, it is the convex hull of the union of all the fences associated with the locations in the itinerary of $a_0$, denoted as $a_0.AccumulatedFence^{S_n}$. A moving actor $a$'s accumulated itinerary since the last snapshot start is $a.iti^{S_n}$= \{$a.l_1^{S_n}$, ..., $a.l_{d-1}^{S_n}$, $a.l_d^{S_n}$\}. The location $a.l_d^{S_n}$ is the latest update location before snapshot $S_n$ updates begin; it is also the first location for next snapshot $S_{n+1}$, shown as $a.l_1^{S_{n+1}}$. The spatial predicate $p$ of $a_0$ is satisfied when $a.iti^{S_n}$ meets $p$ against $a_0.AccumulatedFence^{S_n}$. That is, we check in this example whether $a.iti^{S_n}$ crosses the $a_0.AccumulatedFence^{S_n}$. In summary, the result of $StartReactiveSensing(p, m)$ satisfies Actor-Based Snapshot semantics if the following holds: 

\begin{figure}[!ht]
  \centering
  \includegraphics[scale=0.8]{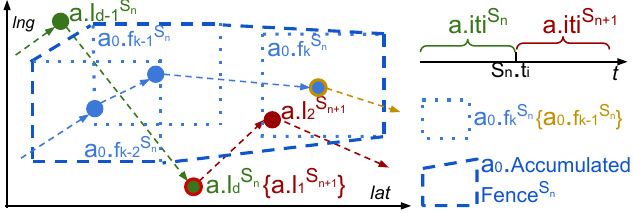}
   \vspace{-0.5cm}
  \caption{Actor-Based Snapshot Semantics for StartReactiveSensing}
  \label{fig:snapshotstartreactivesensing}
  \vspace{-0.5cm}
\end{figure}

\begin{definition}
\label{def:sensing_snap}
The reactive method $m$ in $a_0$ is triggered once at snapshot $S_n$ if and only if $a.itin^{S_n}$ satisfies spatial predicate $p$ against $a_0.AccumulatedFence^{S_n}$. 
\end{definition}

\subsubsection{Actor-Based Freshness semantics vs. Actor-Based Snapshot semantics}

Actor-Based Freshness semantics provides the most recent updates in M-AODBs. However, under Actor-Based Freshness semantics, query processing time affects the amount of uncertainty in terms of location update time
in the results. In general, the longer the query processing time, the more uncertain it becomes when the location updates in the results are carried out wrt. each other. Actor-Based Snapshot semantics is needed when sufficiently close to point-in-time results are needed. Actor-Based Snapshot semantics provides an image of the moving objects' locations where different locations can be contrasted with each other as they are from roughly the same time. However, these self-consistent results are achieved by staling updates. The results from Actor-Based Snapshot semantics could be stale and the construction of the snapshot is expensive, but the generated results are not affected by the query processing time. Therefore, for complex queries that take a long time to process, or queries that can tolerate data staleness but not data inconsistency, snapshot semantics is recommended. For instance, the calculation of object density would likely prefer stale results over a point-in-time snapshot, rather than fresh results over inconsistent location data. Actor-Based Snapshot semantics is also adequate for inter-object distance queries asking for the distance between pairs of objects at a point in time, because returning results with object locations at different time points would result in incorrect distances. In short, the type of semantics should be chosen based on the specific application scenarios, depending on whether the relaxed Actor-Based Freshness semantics is acceptable for the queries and reactions in the application.
  
\section{Dolphin: Design and Implementation} 
\label{sec:dolphin}
\textbf{Dolphin} is our prototype implementation of an M-AODB that supports the moving actor abstraction. We design and implement Dolphin as a library to extend \textit{Microsoft Orleans}~\citep{orleansmicrosoft}, which is a virtual actor programming framework for building robust, distributed systems in the cloud. Orleans provides the high-level programming abstraction of virtual actors while handling the complex responsibilities of actor lifecycle and cluster management. The design of Dolphin maintains all the benefits of Orleans while enabling the programming abstraction of moving actors. 

We chose Orleans because of the maturity of the system and the convenience of the high-level programming model, which makes it a natural starting point for the design of an M-AODB. However, 
the programming abstraction of moving actors is not tightly coupled to Orleans. Our formalization in Sections~\ref{sec:m-aodb} and~\ref{sec:semantics} is carried out on top of the notion of virtual actors, and thus the abstraction would be directly implementable in a framework such as Orbit~\cite{orbit}. Furthermore, we believe that our design is general enough so that it could be adapted with a moderate effort to other actor programming frameworks such as Akka~\cite{akka}.  

\subsection{Moving Actor API in Dolphin}
The actor-oriented database vision~\citep{DBLP:conf/icde/Bernstein18} advocates pluggability of various database features, e.g., transactions, indexing, or streaming, to allow the application to choose and use the required features. Since C\# does not support multiple inheritance using classes, we support pluggability of the moving actor abstraction by exposing it conceptually as a \texttt{mixin} using a C\# \texttt{interface}. An application can define a moving actor by instantiating the moving actor mixin as a property in a virtual actor, thus allowing it to freely compose or inherit any class that it needs to.

{\color{yellow!50!black}\textbf{\textsf{IMovingActorMixin}}} (cf. Listing~\ref{lst:movingactormixin}a) declares the core moving actor operations described in Section~\ref{sec:movingactorform}.  
{\color{cadmiumgreen}\textbf{\textsf{MovingActorMixin}}} (cf. Listing~\ref{lst:movingactormixin}b) is a class that implements {\color{yellow!50!black}\textbf{\textsf{IMovingActorMixin}}} and provides default implementations for all operations in the {\color{yellow!50!black}\textbf{\textsf{IMovingActorMixin}}} interface.\footnote{We could alternatively employ default interface methods in C\#, albeit at the cost of forcing users of Dolphin to commit to C\# 8.0 or later.} These default implementations cover the design and workflows described in Sections~\ref{sec:dolphin:design} and~\ref{sec:query:workflows}.

\begin{lstlisting}[mathescape=true]
public interface IMovingActorMixin {
    Task Move(Point l$_d$);
    Task <List<ActorInfo>> FindActors(Envelope q);
    Task StartReactiveSensing(Predicates p, Func<ReactionInfo,Task> foo);
    Task StopReactiveSensing();
}
\end{lstlisting}

\begin{lstlisting}[mathescape=true, label={lst:movingactormixin}, caption={(a) top: moving actor interface in Dolphin. (b) bottom: default implementations of interface.}]
public class MovingActorMixin : IMovingActorMixin {
    async Task IMovingActorMixin.Move(Point l$_d$){...}
    async Task<List<ActorInfo>> IMovingActorMixin.FindActors(Envelope q){...}
    async Task IMovingActorMixin.StartReactiveSensing(Predicates p, Func<ReactionInfo,Task> foo){...}
    async Task IMovingActorMixin.StopReactiveSensing(){...}
}
\end{lstlisting}

\subsection{Dolphin's Actor-Based Design}
\label{sec:dolphin:design}
To build Dolphin, we implemented the components of an M-AODB outlined in Section \ref{sec:m-aodb:architecture}, namely spatiotemporal concurrency control, spatial indexing, and reactive event monitoring using the Orleans virtual actor programming model. Dolphin is a partitioned system where grid partitioning is employed to logically divide the space into cells. Currently, only uniform grid partitioning is supported in the system where developers can define the size of the cell. However, various spatial partitioning methods, such as learned index structures~\citep{DBLP:conf/sigmod/KraskaBCDP18, DBLP:conf/sigmod/NathanDAK20}, can be employed to better deal with data with skewed spatial distribution. In the current paper, we focus on the design and implementation of the overall system and defer the exploration of spatial partitioning functions to future work. 

Virtual actors (referred to as grains in Orleans\footnote{In the rest of the paper, we use the terms virtual actor, actor, and grain interchangeably unless stated otherwise.}) encapsulate data while providing the abstraction of a lightweight thread that executes methods invoked on it sequentially. Thus, concurrent method invocations on the same actor are processed sequentially while concurrent method executions on different actors happen in parallel. Method invocation on virtual actors is exposed through asynchronous function calls (\texttt{Promises} in C\#), which allows the caller to invoke multiple function calls over multiple actors in parallel. 

\begin{wrapfigure}{l}{0.23\textwidth}
    \includegraphics[width=0.23\textwidth]{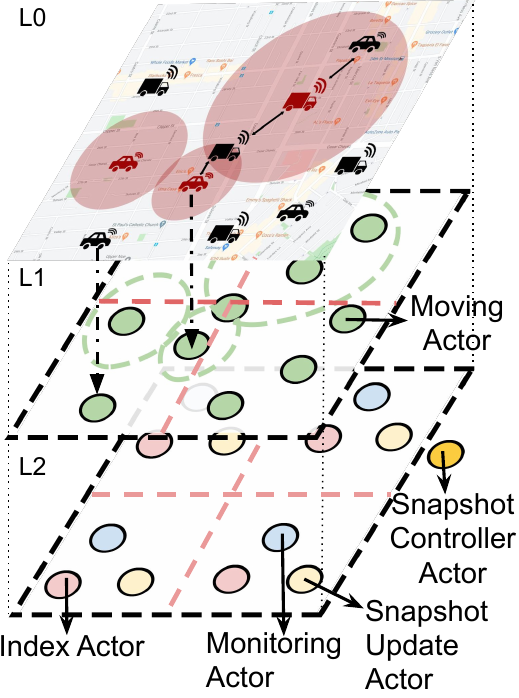}
    \vspace{-.6cm}
    \caption{Dolphin Design: An Example for C-ITS}
    \vspace{-0.5cm}
    \label{fig:dolphindesign}
 \end{wrapfigure}

Figure~\ref{fig:dolphindesign} illustrates the mapping of a representative C-ITS application to various components (actors) in Dolphin's architecture. L0 represents a real-world C-ITS application, where the circles around the vehicle icons represent the sensing fences of physical moving objects. A developer models the digital representation of the physical moving objects using \textit{moving actors} shown as a green circle in L1. The fence of a moving actor is shown as green dashed circles in L1. Space is logically partitioned into uniform grid cells.

L2 shows the functional system components in Dolphin that have been modeled as actors. Each cell consists of an \textit{Indexing Actor} (red circles), \textit{Monitoring Actor} (blue circles),  and \textit{Snapshot Update Actor} (light yellow circles). The \textit{Snapshot Controller Actor} (a deep yellow circle in L2) is not present per cell (placed outside the grid) because it is a single actor for the entire application. The roles of these actors have been outlined below.

\noindent\textit{\textbf{Indexing Actor}}.  It is responsible for indexing locations of moving actors in the cell. 
We currently implement R-trees for indexing the location of moving objects in the system.\\
\noindent\textit{\textbf{Monitoring Actor}}. It is responsible for helping moving actors continuously monitor on their fences against predefined predicates and generate reactions. A monitoring actor acts as an endpoint for communication between moving actors so it receives updates from moving actors in its cell and then relays those updates to all the moving actors who have subscribed to it 
using the Orleans Streams API.\\
\noindent\textit{\textbf{Snapshot Update Actor}}. It is only used to support Snapshot semantics. It is responsible for collecting buffered location updates of moving actors in the cell, dispensing received data to related monitoring actors, and distributing them to other snapshot update actors (if necessary), then finally applying the updates by dispatching them to the indexing actor. Since moving actors may move arbitrarily during a snapshot update interval, the data received by a snapshot update actor may contain location data belonging to other cells, which it needs to relay to the appropriate snapshot update actors.\\
\noindent\textit{\textbf{Snapshot Controller Actor}}. It is only used to support Snapshot semantics. It is responsible for aiding snapshot update actors determine when updates for a snapshot should be reflected on indexes. This coordination is necessary to ensure all the updates for a snapshot have been received from other snapshot update actors before an index is updated in case actors move across cells. It is also necessary in the case where there are no crossings between cells by actors to ensure snapshots across cells do not diverge and are consistent.

We elaborate upon the workflows for processing the operations exposed by the moving actor abstraction utilizing the aforementioned actors in the next section of this paper.

\subsection{Query and Move Workflows in Dolphin}
\label{sec:query:workflows}
\begin{wrapfigure}{l}{0.15\textwidth}
  \vspace{-0.3cm}
  \includegraphics[width=0.14\textwidth]{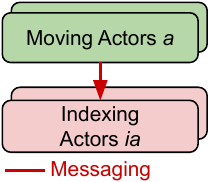}
  \vspace{-0.3cm}
  \caption{Query Workflow}
  \vspace{-0.6cm}
  \label{fig:queryworkflows}
\end{wrapfigure}

In this section, we present the workflow of \texttt{FindActors($r$)} and \texttt{Move($l_d$)} methods of the moving actor abstraction spanning the functional components (actors) in the system under the Actor-Based Freshness and Snapshot semantics.

\subsubsection{Query Workflow under Actor-Based Freshness and Snapshot Semantics}
Figure \ref{fig:queryworkflows} shows the query workflow in Dolphin under both Actor-Based Freshness and Snapshot semantics. When a moving actor $a$ receives a \texttt{FindActors($r$)} request, it first determines the cells that are spanned by the requested query range $r$. Then $a$ asynchronously messages each of the indexing actors $ia$ of those cells to execute an index lookup for the range overlapped by that cell and the requested query range. The communication between the moving actor and the indexing actors is done asynchronously so that the index lookups are performed in parallel by the indexing actors. Note that in snapshot semantics, every index maintains a monotonically increasing version number that is updated when the index is updated at the end of the snapshot to ensure that a query reads data from indices with the same version number. If the version numbers read from the indexing actors are not the same, then the query result is discarded and the query is retried.

\subsubsection{Move Workflow under Actor-Based Freshness Semantics}
\begin{figure}[th!]
  \centering
  \includegraphics[width=0.45\textwidth]{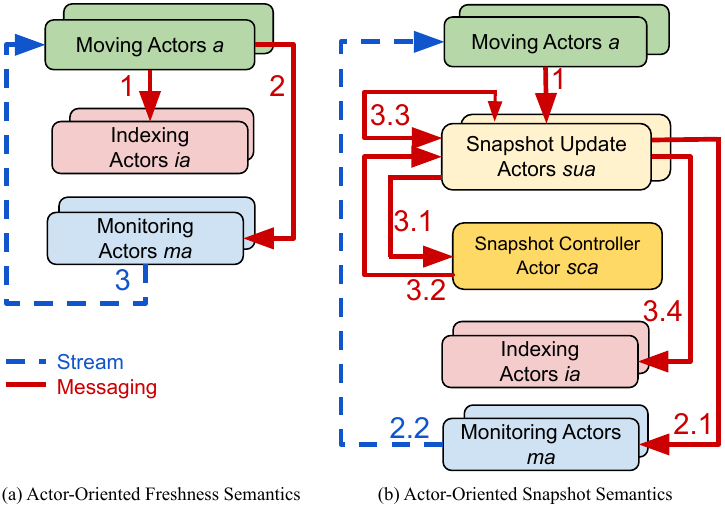}
  \vspace{-0.5cm}
  \caption{Move and Reaction Workflow in Dolphin}
  \label{fig:updateworkflows}
  \vspace{-0.45cm}
\end{figure}

Figure \ref{fig:updateworkflows}~(a) shows the move workflow in Dolphin under Actor-Based Freshness semantics. Since reactions are generated during move, they are explained under the move workflow over the next steps:

\label{sec:move:reaction:workflows}

\noindent
\textbf{\textit{Step 1.}} When a moving actor $a$ receives a \texttt{Move($l_d$)} request, it gets the itinerary ($a.iti=\{l_s, l_d\}$ in Definition \ref{def:findactor_fresh}) and finds the cells spanned by $a.iti$. Then $a$ sends the information of this update to the indexing actors in the cells (one in case of a local move, two in case the actor crosses cells) using asynchronous messages. In the meantime, $a$ also updates its state that consists of the location (from $l_s$ to $l_d$), fence (from $f_s$ to $f_d$), and subscription to the necessary monitoring actors using the Orleans Stream API so that it can receive updates of other moving actors in cells within range of its fence for generating reactions. Note that only moving actors that have enabled sensing using \texttt{StartReactiveSensing($p$,$m$)} subscribe to the stream.

\noindent
\textbf{\textit{Step 2.}} Moving actor $a$ sends the update information to the monitoring actors spanned by $a.iti$ so that necessary reactions can be generated on the moving actors that have sensing enabled.

\noindent
\textbf{\textit{Step 3.}} When a monitoring actor $ma$ receives updates from moving actors, it publishes the updates on its stream channel\footnote{Every cell has a single stream channel with the monitoring actor being the sole publisher and moving actors with sensing enabled being the subscribers.} that is consumed by the moving actors. Recall from Definition \ref{def:sensing_fresh}, only a moving actor (e.g., $a_0$) that has enabled sensing by invoking \texttt{StartReactiveSensing($p$,$m$)} receives location updates from the monitoring actors on the subscribed stream channel, then checks if the received $a.iti$ satisfies its spatial predicate $p$ against latest fence $f$ (either $f_s$ or $f_d$, depending on the relation between $a.t_u$ and $a_0.t_u$ shown in Definition \ref{def:sensing_fresh}). If the spatial predicate is satisfied, then the registered method $m$ is executed to perform the reactive action.

\subsubsection{Move Workflow under Actor-Based Snapshot Semantics}
Figure \ref{fig:updateworkflows}~(b) shows the move workflow in Dolphin under Actor-Based Snapshot semantics. Since reactions are also generated during move, they are explained under the move workflow over the next steps:

\noindent
\textbf{\textit{Step 1.}} When a moving actor $a$ receives a \texttt{Move($l_d$)} request, it updates its state that consists of its location and fence, and buffers the updated state locally. Aligned with Definition \ref{def:sensing_snap}, this process creates $a.itin^{S_n}$ for any moving actor and $a_0.AccumulatedFence^{S_n}$ for moving actors who have started sensing. When the timer in the moving actor\footnote{Every moving actor uses a timer configured to the snapshot interval time that fires periodically to signal the start of a snapshot.} expires at $S_n.t_i$ mentioned in Definition \ref{def:findactors_snap}, $a$ finds the cell that the first buffered location refers to, and sends the entire buffered itinerary $a.itin^{S_n}$ to the snapshot update actor $sua$ of this cell. Later data exchanges for snapshot consistency and reaction correctness are carried out between snapshot update actors. 

\noindent
\textbf{\textit{Step 2.1.}} Every snapshot update actor $sua$ tracks the moving actors that are present in its cell, which is updated at the end of every snapshot. When a snapshot update actor $sua$ receives the buffered location data from all the moving actors that were present in its cell in the last snapshot, it sends the buffered itineraries to monitoring actors $ma$ in the cells that spanned those itineraries. 

\noindent
\textbf{\textit{Step 2.2.}} When a monitoring actor $ma$ receives the buffered itineraries, it distributes them to the sensing moving actors using the mechanism (Step 3) outlined in the move workflow under Actor-Based Freshness semantics. To fulfill Definition \ref{def:sensing_snap}, when a sensing moving actor $a_0$ receives itineraries $a.itin^{S_n}$, it checks them using its predicate $p$ against  $a_0.AccumulatedFence^{S_n}$.

\noindent
\textbf{\textit{Step 3.1.}} The following steps help switch from a snapshot $S_{n-1}$ to snapshot $S_n$ (see Definition \ref{def:findactors_snap}). When a snapshot update actor $sua$ receives the buffered data from all the moving actors in its cell in the previous snapshot, it sends a message to the snapshot controller actor $sca$ informing the snapshot update actors they would be sending their buffered location updates to. Recall that there is a single snapshot controller actor in the system.

\noindent
\textbf{\textit{Step 3.2.}} The snapshot controller $sca$ tracks cells (active) that have moving actors in it.  When the snapshot controller actor $sca$ receives the requests from $sua$ in all active cells from Step 3.1, it responds by informing them about the corresponding $sua$s from whom they should receive messages (final location) from.

\noindent
\textbf{\textit{Step 3.3.}} The snapshot update actors $sua$ exchange information between each other to get the needed update data in this snapshot round.

\noindent
\textbf{\textit{Step 3.4.}} After sending all data out and receiving all data from the others, every $sua$ sends update information to its cell's indexing actor to update the indices. After all the indices of all the cells are updated, a new snapshot is completed. 


\section{Experimental Evaluation}
\label{sec:experimentalevaluation}

This section presents experimental evaluation to show that \textit{Dolphin} has met the requirements for reactive moving object data management in Section~\ref{sec:reqs}. Our goals in evaluation are the following:

\vspace{-0.25ex}
\begin{enumerate}[leftmargin=*]
    \item Characterize the relationship between moves and reactions, including aspects such as client-side load and reactive sensing intensity (cf. Sections~\ref{exp:workload},~\ref{exp:reactive_sensing} and~\ref{exp:excessive_reative_sensing});
    \item Observe how spatial range queries affect Dolphin and their interaction with moves and reactions (cf. Section~\ref{exp:queryratio});
    \item Evaluate if Dolphin scales out over multiple machines in a distributed setting (cf. Section~\ref{exp:scaleout});
    \item Investigate how the performance of moves and reactions is affected by spatial skew and realistic spatial distributions (cf. Sections~\ref{exp:skew_throughput} and~\ref{exp:c-its}).
\end{enumerate}
\vspace{-1ex}


\subsection{Experimental Setup}
\subsubsection{Cloud Service and Deployment}

We deployed several components for the experiments. The first one was Orleans silo \citep{orleanssilo}, where virtual actors are activated in a server and run all of the application logic. We set up our benchmark environment on AWS~\citep{aws}. We employed Amazon DynamoDB \citep{dynamodb} for storing the silo membership table \citep{clustermanagement} and we utilized EC2~\citep{ec2} for deploying all our instances. Three types of EC2 computing optimised instances were employed. One c5.xlarge instance for controlling and synchronising benchmark client threads; and one c5.4xlarge instance to simulate moving object client threads. Eight c5.xlarge instances are used as silos to run distributed experiments; and one c5.9xlarge instance to run single-silo experiments. All instances run on Windows Server 2019 Base and Orleans 3.1.2. We placed them in one subnet and one cluster deployment group to reduce network latency \citep{DBLP:conf/cloud/ZouWSBDGW11}. We built a spatial-preference grain placement strategy \citep{GrainPlacement} to reduce the number of messages across silos (cf. Subsection \ref{exp:scaleout}). 

\subsubsection{Query, Move, and Reactions}
Our experiments start by initialising a set of moving actors. All clients issue a sequence of requests as soon as possible. Client requests only include queries and moves to moving actors. We define the following processes in Dolphin:

\minisec{Query} Queries are range queries 
using the method \texttt{FindActors(r)}. In our experiment, we fix the query window size for simplicity. 

\minisec{Move} Moves are 
\texttt{Move($l_d$)} requests. 
Moves may trigger a reactive method of all moving objects that have started reactive sensing accordingly.  
Client benchmarks were responsible for interpreting movement models and generating the necessary move calls.

\minisec{Reaction}
A fraction (0-100\%) of moving actors are set to call \texttt{StartReactiveSensing($p$, $m$)} at initialization so that each triggers reactive action $m$ when other actors' move requests satisfy $p$ against its fence. In our experiments, for all moving actors, we chose \emph{cross} as a spatial predicate. We also defined the reactive method as building a connection with the actor who crosses the fence. 

This section analyses performance issues concerning underlying workflows associated with \emph{queries} and \emph{move}. We called them client requests as they are triggered by clients. Differently, reactions themselves occur when reactive sensing is on.
Related work on moving objects considers two kinds of requests: queries and move, but without reactions. Our work better reflects the real world as we include reactive behaviors; as such, our experiments cannot be directly comparable to experiments performed in related work. On the other hand, reactions come with a processing cost.
Therefore, while in other systems, the \emph{move} request does not imply additional processing, moves in Dolphin are more expensive than in other works due to associated reaction processing.

\subsubsection{Benchmark Implementation}
\label{benchmark}
\begin{figure}
 \vspace{-0.5cm}
    \begin{minipage}{0.15\textwidth}
        \centering
        \includegraphics[width=0.8\linewidth]{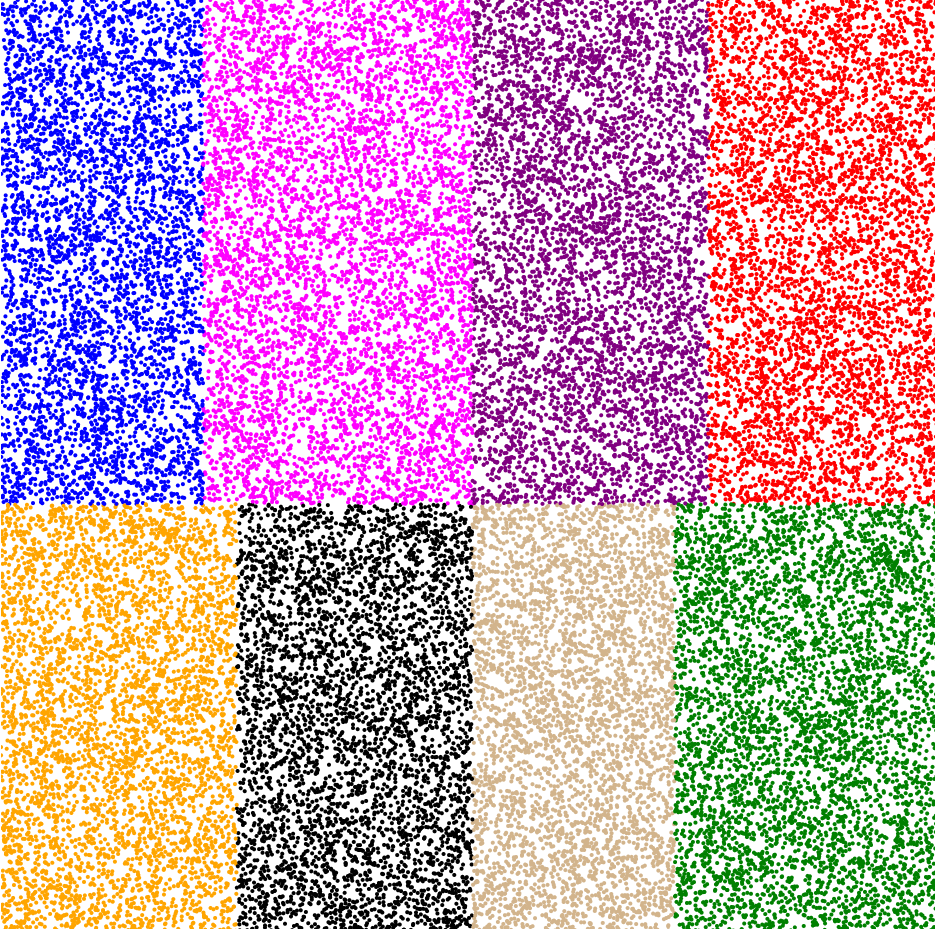}
        \vspace{-0.3cm}
        \caption{uniform distribution benchmark dataset.} 
        \label{fig:uniformdataset}
    \end{minipage}\hfill
    \begin{minipage}{0.15\textwidth}
        \centering
        \includegraphics[width=0.8\linewidth]{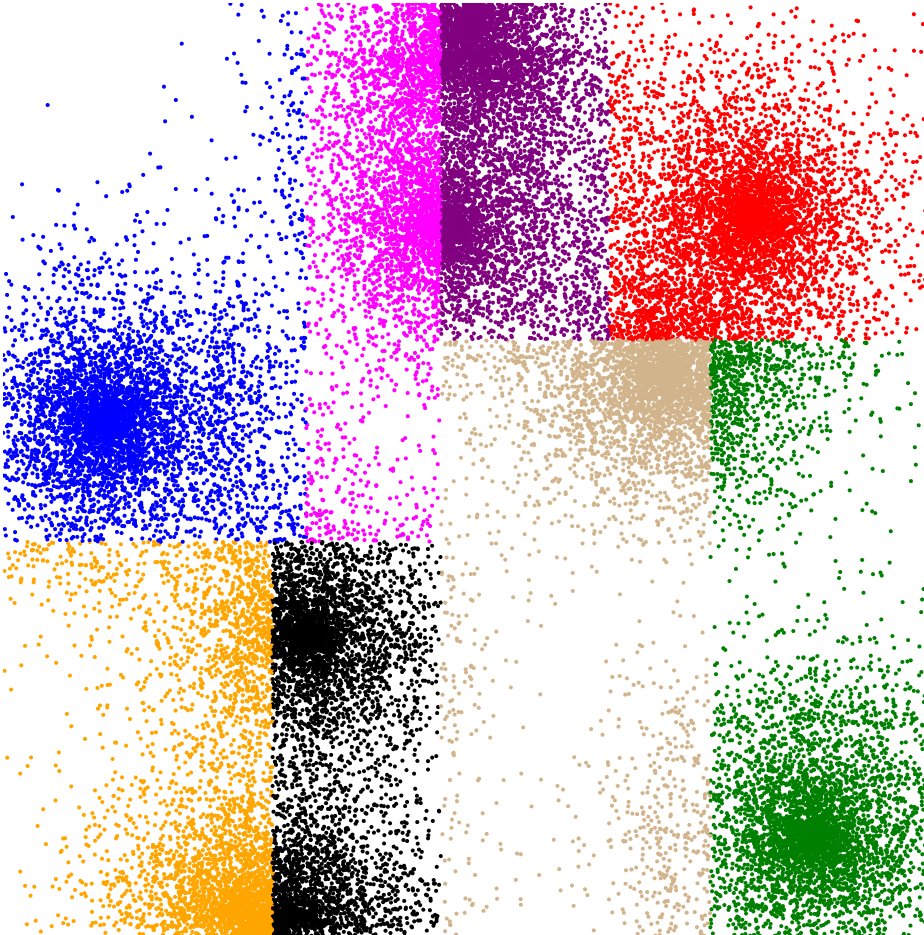}
         \vspace{-0.3cm}
        \caption{gaussian distribution benchmark dataset.} 
        \label{fig:gaussiandataset}
   \end{minipage}\hfill
     \begin{minipage}{0.15\textwidth}
       \vspace{0.3cm}
        \includegraphics[width=1\linewidth]{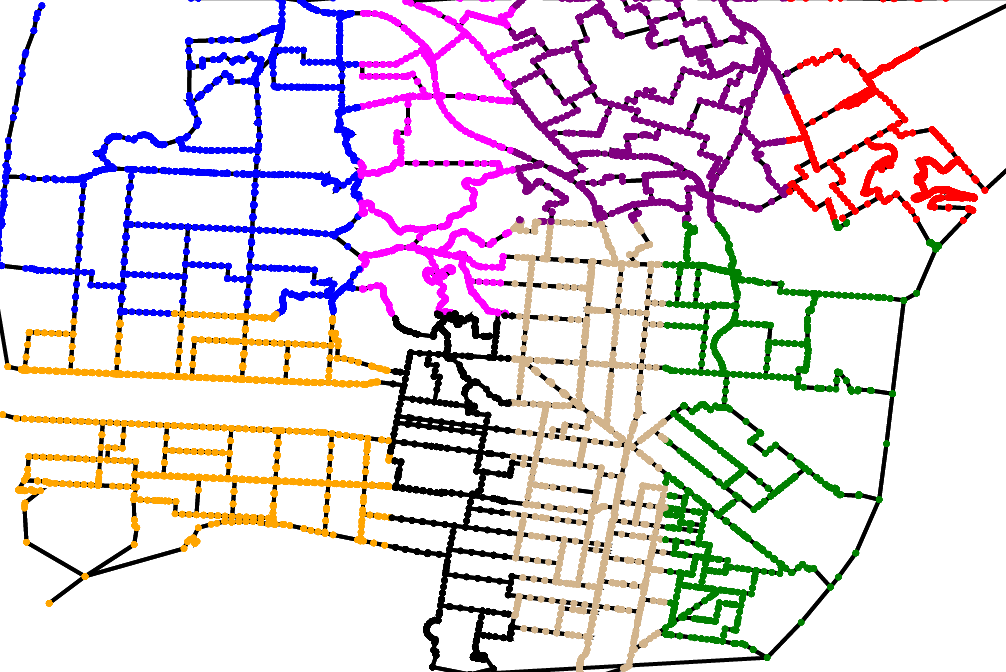}
        \vspace{-0.6cm}
        \caption{c-its scenario benchmark dataset.} 
        \label{fig:citsdataset}
    \end{minipage}
     \vspace{-0.5cm}
\end{figure}

We conducted experiments under three benchmarks, namely uniform distribution benchmark (cf. Figure \ref{fig:uniformdataset}), Gaussian distribution benchmark (cf. Figure \ref{fig:gaussiandataset}) 
and C-ITS scenario benchmark (cf. Figure \ref{fig:citsdataset})
. 
In the first two benchmarks, 
moving objects can move freely in a 2D space under two distributions -- uniform or Gaussian. In the C-ITS scenario benchmark, objects are constrained to move along a real-life urban network. We set the configuration of our benchmarks from \citep{DBLP:journals/pvldb/ChenJL08, DBLP:journals/pvldb/SowellSCDG13}, adapting their settings to our scenario (cf. details at Subsection \ref{workloads}).

\minisec{Uniform Distribution Benchmark}
In the uniformly distributed dataset, moving actors are instantiated uniformly at random in positions in a fixed size square space and move towards random directions with a random speed under a speed limitation.

\minisec{Gaussian Distribution Benchmark}
The Gaussian distributed dataset models a skewed spatial workload. 
In this configuration, moving actors are clustered around a set of hotspots. Those hotspots are uniformly initialised at random locations, and moving actors are distributed around each hotspot based on a Gaussian distribution. Actors move at a speed related to the location of the hotspot. A moving actor moves faster when it is farther away from hotspots and takes a slower movement when close to hotspots~\citep{DBLP:journals/pvldb/ChenJL08, DBLP:journals/pvldb/SowellSCDG13}. 

For both uniformly and Gaussian distributed datasets, if a moving actor moves across the borders of the square space, it is bounced back to the opposite direction and moves the same excess distance that should not exceed the border. If this is not possible, the moving actor remains stationary and expects a new update. 

\minisec{C-ITS Scenario Benchmark}
The C-ITS scenario benchmark models constrained spatial movement with the aim of observing the performance of Dolphin in a realistic setting. For this benchmark, we used the TIGER/Line data \citep{tigerline} corresponding to ``All roads2019 San Francisco County". Moving actors behave as digital twins for vehicles in San Francisco. According to the authors of \citep{mobility2018sanfrancisco}, we initialized 38000 vehicles\footnote{5700 transportation network companies (TNCs) vehicles operate during peak period, and those trips represent 15\% of all intra-San Francisco vehicle trips. For the sake of simplicity, we estimate 5700/15\%=38000, which may be less than real traffic because more commute trips exist in peak period.} and limited the space to the main part of San Francisco city. 
Furthermore, vehicles move along the road network at a fixed speed (22m/s \citep{carspeedlimits}), while following two rules: (1) A moving actor keeps moving along an edge; and (2) when the moving actor arrives at the conjunction of edges, it chooses to follow a random direction along the edges and continues.

\subsubsection{Workloads}
\label{workloads}

Table \ref{tab:workload} presents our workload settings, where parameters are adapted whenever possible from previous benchmarks~\citep{DBLP:journals/pvldb/SowellSCDG13, DBLP:journals/pvldb/ChenJL08}. To evaluate our system under high, but still realistic, intensity, we deviate from previous settings in the following. We set a maximum speed of moving actors to be 80km/h, which is the highest recorded speed limit for urban areas among most countries~\citep{carspeedlimits}. Additionally, we adopted the proportion between moving object numbers and space sizes from the study in \citep{DBLP:journals/pvldb/SowellSCDG13}, since it is also focused in in-memory evaluation. However, in contrast to previous studies that focus on notions of virtual time, \textit{e.g.}, ticks or timestamps \citep{DBLP:journals/pvldb/ChenJL08, DBLP:journals/pvldb/SowellSCDG13}, our study configured the base number of moving actors for one silo to 5,000. We aimed to match a workload that could be easily handled by a silo while delivering reasonable real-time performance. 

\begin{table}
\vspace{-0.3cm}
\captionof{table}{Workloads for Benchmarks.}
\vspace{-0.3cm}
  \label{tab:workload}
  \footnotesize
  \begin{tabular}
  {|@{\hspace{0.04cm}}c@{\hspace{0.04cm}}|@{\hspace{0.04cm}}c@{\hspace{0.04cm}}|@{\hspace{0.04cm}}c@{\hspace{0.04cm}}|@{\hspace{0.04cm}}c@{\hspace{0.04cm}}| }
  \hline
    \multirow{2}{*}{\textit{Parameters}}        & \multicolumn{3}{c|}{\textbf{Benchmarks}}                         \\\cline{2-4} 
                                                & \textbf{Uniform}           & \textbf{Gaussian}       & \textbf{C-ITS}   \\ \hline
    \textit{Number of Servers}                  & 1, 2, 4, 8                 & 8                       & 8               \\\hline
    \textit{Client Threads per Server}          & \begin{tabular}[c]{@{}c@{}}8 (single server) or \\4 (multiple servers) \end{tabular}                       & 4                       & 4               \\\hline
    \textit{Number of Moving Actors}            & 5000 per server              & 40000                   & 38000           \\ \hline
    \textit{Space Size ($km^2$)}                & 100,199.94,400,799.98  & 799.98                  & 154           \\ \hline
    \textit{Number of Cells}                    & 100,196,400,784            & 784                     & 805             \\ \hline
    \textit{Number of Hotspots}                 & /                          & 8,80,400,800,8000,40000 & /               \\ \hline
    \textit{Fence Size ($m^2$)}                 & \multicolumn{3}{c|}{$1000\times1000$}                                              \\ \hline
    \textit{Query Size ($m^2$)}                 & \multicolumn{3}{c|}{$1000\times1000$}                                              \\ \hline
    \textit{Max Speed ($km/h$)}                 & \multicolumn{3}{c|}{80}                                                \\ \hline
    \textit{Reactive Sensing Percentage ($\%$)} & 0,\textbf{12.5},25,50,100    &12.5                     &12.5             \\ \hline
    \textit{Query Ratio ($\%$)}                 & \textbf{0},20,40,60,80,100 & 0                       & 0               \\ \hline
    \textit{Snapshot Interval Time ($s$)}                & \multicolumn{3}{c|}{1 or 4}                                            \\ \hline
  \end{tabular}
\vspace{-0.5cm}
\end{table}

In line with previous studies, however, the fence and query sizes were chosen based on a realistic setting, where vehicles are interested in a range of 1km. We set the default query rate to be 0, which means all client's requests are move requests to maximize the reaction influence. There is no guide for reactive sensing percentage setting in previous studies. We then set and assume it to be 12.5\% based on a reality assumption. We set the number of cells in the spatial partitioning scheme used (100) as well as the snapshot interval time (1~sec) based on corresponding tuning experiments. Based on snapshot interval time tuning experiments, for a single server, the system can achieve snapshot update within 1 sec. Also, the system provides a reasonable reaction throughput and 1 sec can satisfy most use cases. The snapshot interval time for distributed experiments are set to be both 1 sec and 4 secs. 

In our setup, c5.xlarge instances were used for the Orleans silo in most experiments except a experiment shows in Section~\ref{exp:excessive_reative_sensing}. Every client thread was configured to send requests as fast as possible to moving objects so that we can increase update frequency and thus trigger reactions more quickly to the application. Our focus was on the scalability of updates and reactions, which stands in contrast to previous studies, where each moving object only sends updates infrequently. In previous studies, scalability is sought in terms of the numbers of moving objects~\citep{DBLP:journals/pvldb/ChenJL08}. 
Based on our workload tuning experiments results (cf. Figure \ref{fig:workload}), clients only send move requests because we emphasized exploring reaction behavior. When client threads are 8, the workload saturates our 4vCPUs server. Meanwhile, moving actors update frequency is around every 0.6 secs and every 0.5 secs under Actor-Based Freshness and Actor-Based Snapshot semantics, respectively. For distributed experiments, we choose not to saturate each server to give some resources to extra overhead. In this sense, we used four client threads for each server. 

\subsection{Experimental Results}

This section presents the experimental results. For brevity, we use \textsf{Fresh} and \textsf{Snap} to express 'Actor-Based Freshness semantics' and 'Actor-Based Snapshot semantics', respectively.

\subsubsection{How do reactions behave as Dolphin is faced with increasing move workloads?}
\label{exp:workload}

\begin{figure*}[!t]
    \begin{minipage}{0.30\textwidth}
        \centerline{\includegraphics[width=1\linewidth]{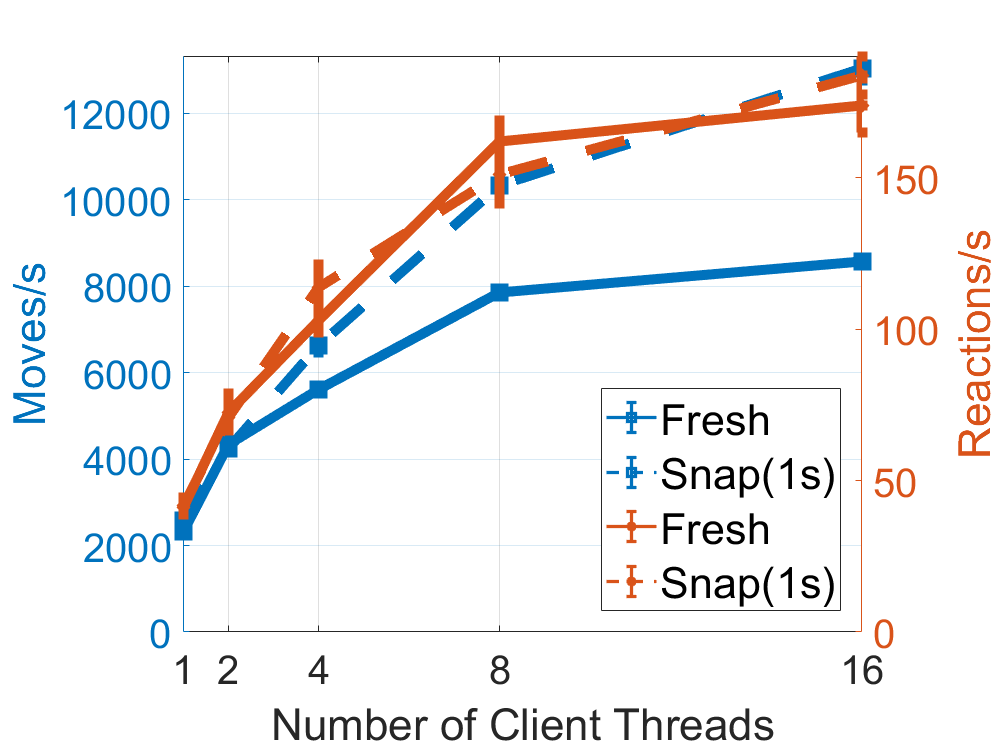}}
        \vspace{-2ex}
       \caption{Correlation of reaction and move throughputs under increasing client-side workload intensity.} 
        \label{fig:workload}
  \end{minipage} \hfill
    \begin{minipage}{0.30\textwidth}
        \centerline{\includegraphics[width=1\linewidth]{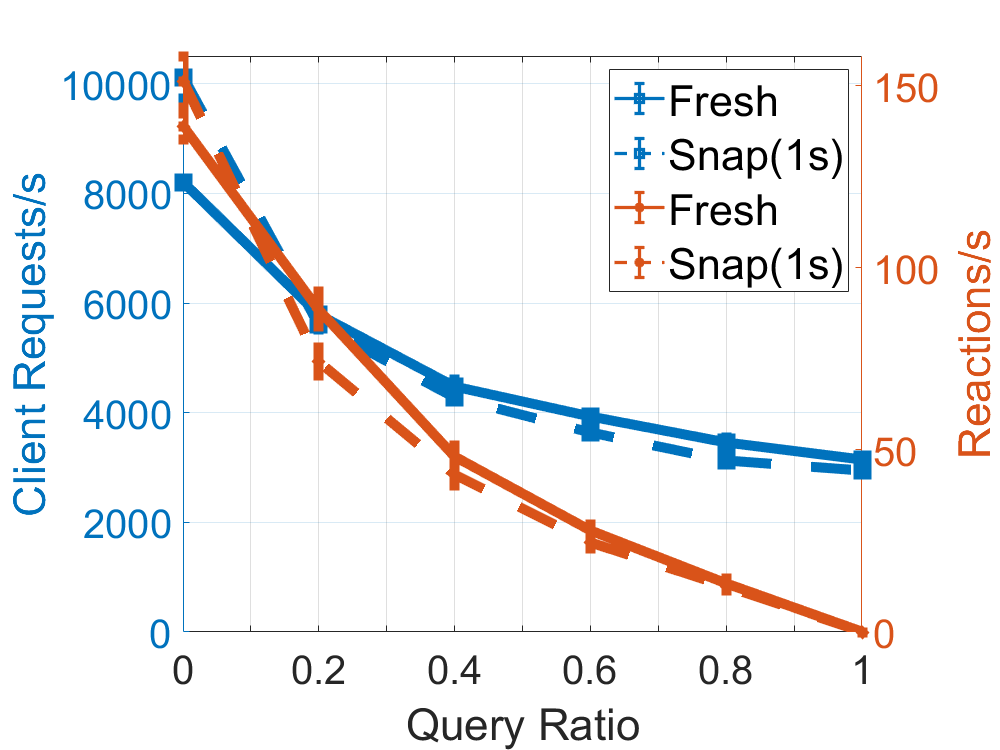}}
        \vspace{-2ex}
        \caption{Client request and reaction throughputs under increasing query ratio.} 
        \label{fig:queryrateT}
   \end{minipage} \hfill
    \begin{minipage}{0.30\textwidth}
        \centerline{\includegraphics[width=1\linewidth]{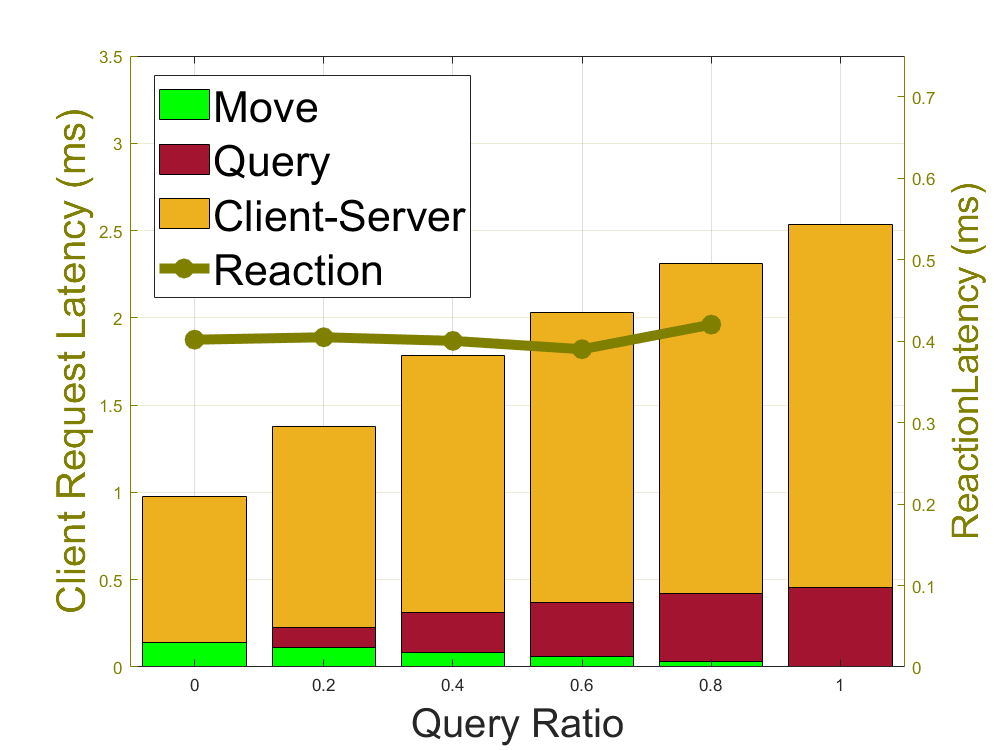}}
        \vspace{-2ex}
        \caption{Latency breakdown with increasing query ratio under Freshness semantics.} 
        \label{fig:queryrateFL}
    \end{minipage}\hfill
    \vspace{-0.5cm}
\end{figure*}

In this experiment, we increased the client-generated move workload up to the saturation point of a single server, and observed the effects on reactions. Figure \ref{fig:workload} shows that reaction throughput increases along with move throughput for both \textsf{Fresh} and \textsf{Snap(1s)}. This is expected because more moves increase the chances to trigger more reactions. However, reaction generation also increases the resource consumption on the server side, which limits the attainable throughput of moves. Since \textsf{Snap(1s)} batches reaction generation at the snapshot interval of 1s, its move throughput is more resilient to interference and is 1.52x higher than \textsf{Fresh} at 16 client threads. We found that 8 client-thread workloads saturate the system well. With this setting, we observe move request latencies of 1.02~ms and 0.77~ms measured at the client side for \textsf{Fresh} and \textsf{Snap(1s)}, respectively. Since there are 5000 moving actors in a server, each moving actor is able to report a move on average every 0.637~sec or 0.484~sec depending on the semantics. 
This result shows that our system can handle most cases in real life.

\subsubsection{How does increasing spatial range query ratio affect Dolphin's request and reaction throughputs and latencies?}
\label{exp:queryratio}

Except for the new reactive feature, Dolphin provides range query and moves like other spatial data management systems. We have explored how reactions and moves affect the system. We now proceed to describe the performance of Dolphin regarding different kinds of client requests. This scenario is a mix of ``query plus move" in different query ratios in order to understand trade-offs in workload configuration and overall performance costs. Query ratio is the percentage of client query requests against all client requests. We varied query ratio from 0\% to 100\%, which means varied client requests from composed by 100\% move, 0\% query to 100\% query, 0\% move. Meanwhile, 12.5\% moving actors are reactive moving actors in this experiment to understand how query ratio affects the reaction generation. 

Figure \ref{fig:queryrateT} shows that for both \textsf{Fresh} and \textsf{Snap(1s)}, as expected, reaction throughput decreased with increased query ratio because reactions are generated by move. Client request throughput also goes down along with increased query ratio. This effect can be analyzed by the interplay of query and reaction expenses and a breakdown of the client request latency under \textsf{Fresh}. Our results in the latency analysis of Figure~\ref{fig:queryrateFL} show that a query is about 3.44x-4.10x more expensive than a move in this scenario. However, client request throughput only degrades by 2.60x. 
That is explained by reactions being more relatively expensive in Dolphin. With the decrease of moves, the influence of reactions decreased. Therefore, with the increase of query ratio, the degradation in client task throughput is not so dramatic in Dolphin. A similar result is found in \textsf{Snap(1s)} as well.

\subsubsection{Can Dolphin scale out move and reaction throughputs with increasing servers and workloads?}
\label{exp:scaleout}

\begin{figure*}[!t]
   \begin{minipage}{0.30\textwidth}
        \centerline{\includegraphics[width=1\linewidth]{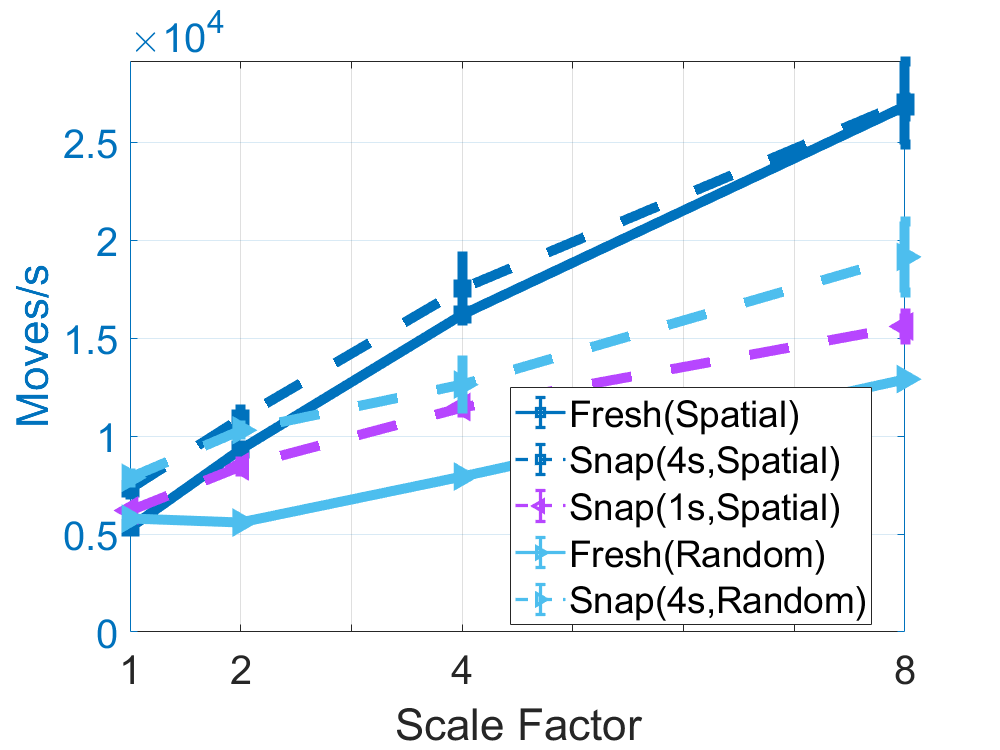}}
         \vspace{-2ex}
        \caption{Move throughput scalability with increasing servers and workloads.} 
        \label{fig:scaleoutT}
   \end{minipage}\hfill
    \begin{minipage}{0.30\textwidth}
        \centerline{\includegraphics[width=1\linewidth]{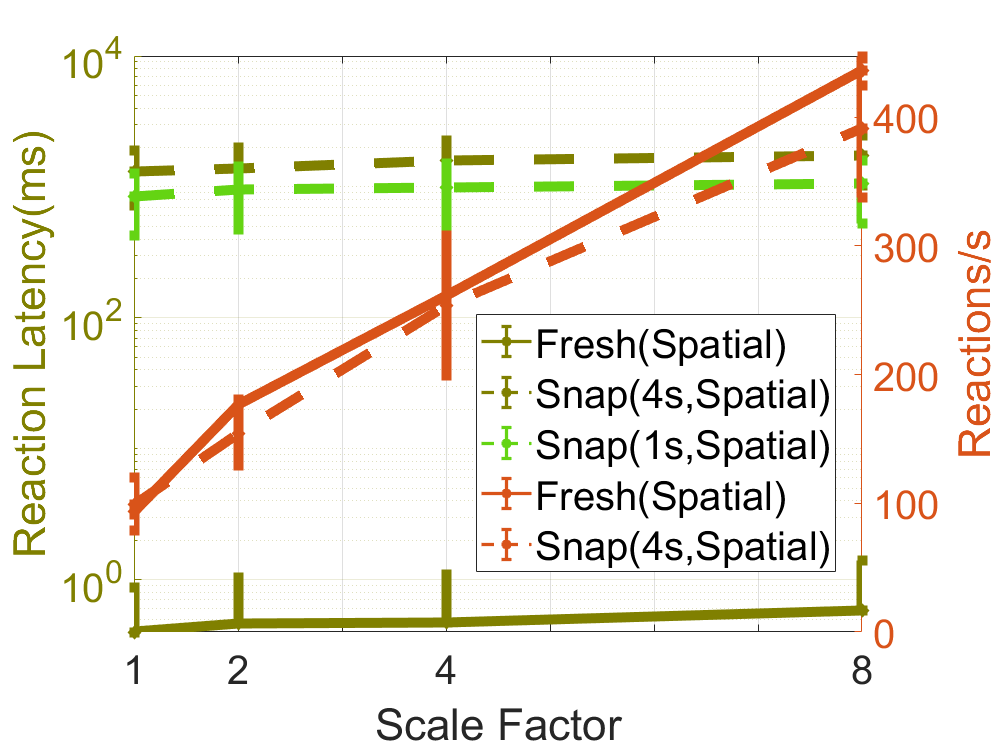}}
         \vspace{-2ex}
        \caption{Reaction latency and throughput with increasing servers and workloads.}
        \label{fig:scaleoutR}
    \end{minipage}\hfill
    \begin{minipage}{0.30\textwidth}
        \centerline{\includegraphics[width=1\linewidth]{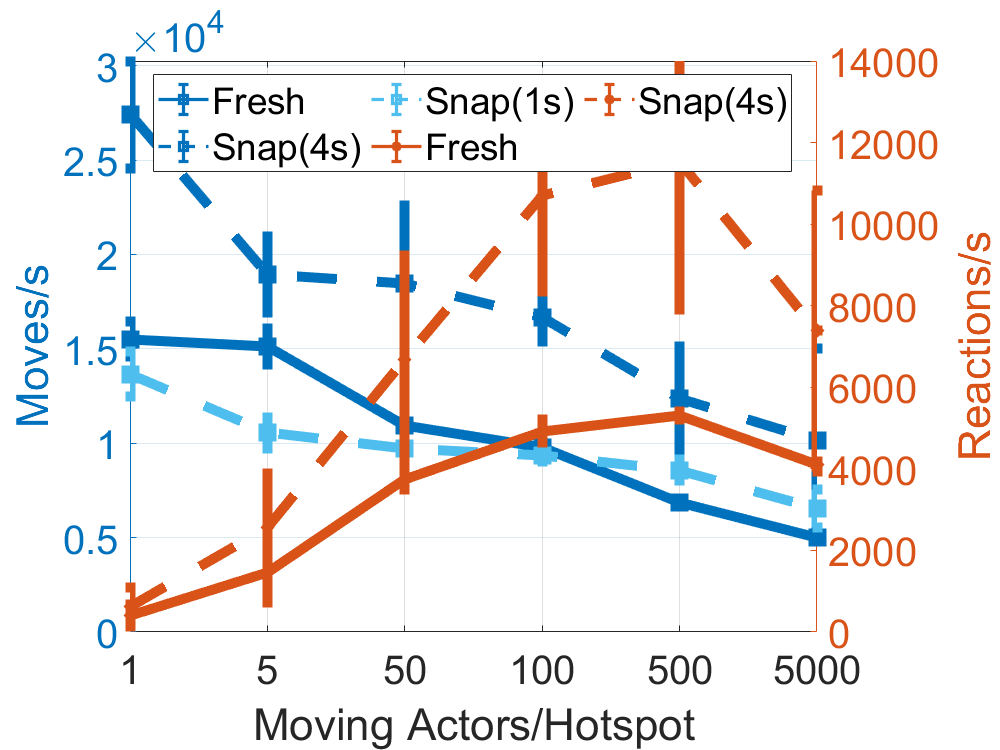}}
        \vspace{-2ex}
        \caption{Move and reaction throughputs with increasing spatial skew.}
        \label{fig:skewT}\hfill
    \end{minipage}
\vspace{-3ex}
\end{figure*}

To test the scalability of Dolphin, we scale the number of servers, datasets, and workloads (cf. Table~\ref{tab:workload}). All client requests are move requests. In Orleans, the actor runtime decides which server to activate an actor (grain) on, which is called grain placement. The default grain placement in Orleans is to activate grains on a random server in the cluster \citep{GrainPlacement}. In Figure \ref{fig:scaleoutT}, move throughput \textsf{Fresh(Random)}, \textsf{Snap(4s, Random)} increases non-linearly afterward because communications among servers increase with the number of servers. To solve that, we developed a customized spatial grain placement that partitioned the whole space by employing a KD-tree~\citep{DBLP:journals/ipl/Orenstein82}. Spatially close moving actors having a higher chance to communicate with each other can then be placed on a server. 

Move throughput scalability under spatial grain placement, especially of \textsf{Fresh(Spatial)}, improved substantially. However, throughput is still not perfectly linearly increased because communication between servers cannot be eliminated, and moving actors may move outside the placement partitions along with time. Moreover, due to the single component to control snapshot update, the throughput of \textsf{Snap(1s, Spatial)} is still not optimal. That can be solved by enlarging the snapshot interval time to reduce the overhead of the snapshot update, seen from the move throughput \textsf{Snap(4s, Spatial)}. We employ our customized spatial grain placement for the following experiments and omit that in legends.

In Figure \ref{fig:scaleoutR}, as the system scales, reaction throughput \textsf{Fresh(Spatial)} and \textsf{Snap(4s, Spatial)} increase along the same trend as the corresponding move throughput in Figure \ref{fig:scaleoutT}. For \textsf{Snap} variants, reaction latency is constrained by snapshot interval time because reactions are only triggered during the snapshot update. For two different snapshot interval times, 50th-percentile reaction latency from 1 sec interval time is 1.53x-1.73x faster than from 4 secs. But they are both uncompetitive to the reaction latency of \textsf{Fresh}, which provides a more than one thousand times faster reaction. So for applications extremely sensitive to real-time reactions, Actor-Based Freshness semantics is recommended.

    
\subsubsection{How do move and reaction throughputs behave under increasing spatial skew?}
\label{exp:skew_throughput}

We employ a Gaussian distributed workload (cf. Table \ref{tab:workload}) to answer this question. All client requests are move requests. We analyzed the move and reaction throughput behavior with increasing spatial skew in data. In Figure \ref{fig:skewT}, as expected, move throughput goes down with more spatial skew, because skewed data increases the workload to a single cell, which causes more queuing and communication in those over-loaded cells. 

We can also see that \textsf{Snap(4s)} provides a better move throughput performance compared with \textsf{Snap(1s)}. For both \textsf{Fresh} and \textsf{Snap(4s)}, reaction throughput increased from near-uniform data (1 moving actor/hotspot) to 500 moving actors/hotspot data, but then decreased for further skew. The reason for increasing reaction throughput is that due to the fixed number of moving actors, with fewer hotspots, more moving actors gather together around one hotspot. In this sense, one single move trajectory is more likely to satisfy predicates against more fences. However, reactions should also decrease along with the decrease of move throughput as shown in Subsection~\ref{exp:workload}. Also, in our Gaussian movement setting, a moving actor in more skewed data would have a lower chance to move faster, which would decrease the length of a trajectory. Those counterfactors make the reaction throughput flatten and finally decrease as in Figure \ref{fig:skewT}.




\subsubsection{How does Dolphin perform in the C-ITS scenario benchmark?}
\label{exp:c-its}

\begin{table}
\small
\captionof{table}{Results for C-ITS scenario benchmark.}
\vspace{-0.3cm}
\label{tab:roadnetwork}
\begin{tabular}
{|@{\hspace{0.04cm}}l@{\hspace{0.04cm}}|@{\hspace{0.04cm}}l@{\hspace{0.04cm}}|@{\hspace{0.04cm}}l@{\hspace{0.04cm}}|@{\hspace{0.04cm}}l@{\hspace{0.04cm}}|}
\hline
 \multirow{2}{*}{\textit{Results}}        & \multicolumn{3}{c|}{\textbf{Semantics}}                         \\\cline{2-4} 
                & \texttt{Fresh}  & \texttt{Snap(1s)} & \texttt{Snap(4s)}      \\ \hline
\textit{Moves/s}       & 3349.99($\pm$34.65)  & 5211.61($\pm$2583.67)  & 9276.87($\pm$2583.67)      \\ \hline
\textit{50\% Latency (ms)} & 6.26               & 0.59                 & 0.55                      \\ \hline
\textit{99\% Latency (ms)} & 53.55              & 89.94                & 60.84                     \\ \hline
\textit{Reactions/s}   & 1925.86($\pm$49.31)  & 120.29($\pm$32.54)     & 287.86($\pm$134.29)         \\ \hline
\textit{50\% Latency (ms)} & 22.56              & 5605.49              & 8261.29                   \\ \hline
\textit{99\% Latency (ms)} & 1092.66            & 18835.36             & 22364.91                  \\ \hline
\end{tabular}
\vspace{-0.4cm}
\end{table}

We generate workload according to the C-ITS scenario (cf. Table \ref{tab:workload}) to understand how Dolphin behaves in a realistic benchmark. Table \ref{tab:roadnetwork} shows the results. We observe that the 50\% latency and 99\% latency of vehicle move under \textsf{Fresh} is 6.26 ms and 53.55 ms, respectively. The corresponding latencies under \textsf{Snap(1s)} is 0.59 ms and 89.94 ms, respectively; and 0.55 ms and 60.84 ms, respectively, for \textsf{Snap(4s)}. According to ETSI ITS Specifications 2020 \citep{etsi}, the threshold for waiting to be serviced by a GNSS Positioning Correction service of a Roadside ITS is 2 seconds. So the move latency provided by Dolphin in the C-ITS scenario benchmark satisfies this requirement well. Also, we note that the 50\% reaction latency of \textsf{Fresh} is 22.56 ms, which provides a near-real-time reaction latency. \textsf{Snap(4s)} provides faster reaction and higher move and reaction throughput than \textsf{Snap(1s)}. Therefore, 4 secs is a more suitable snapshot interval time for our C-ITS scenario benchmark. 




\section{Related work and discussion}\label{sec:relatedwork}

As modern interactive and data-intensive applications demand a scalable and elastic application tier~\citep{bernstein2014orleans}, actor programming frameworks such as Akka~\citep{akka}, Erlang~\citep{erlang}, Orbit~\citep{orbit}, and Orleans~\citep{orleansmicrosoft} are becoming popular implementation options that have led to significantly increased programmer productivity. In particular, the abstraction of virtual actors~\citep{DBLP:conf/cloud/BykovGKLPT11} in Orleans considers actors as modular and stateful virtual entities in perpetual existence, facilitating a one-to-one mapping to moving objects, among other IoT scenarios. The Orleans runtime increases productivity and convenience for developers by handling failure, automatically and dynamically managing actors' life cycle and balancing load across servers, which also helps developers achieve the desired scalability and availability~\citep{DBLP:conf/icde/Bernstein18, bernstein2014orleans}. Orleans has been used in many production services, which validate its readiness and reliability. 

By building upon virtual actors and Orleans, Dolphin leverages many of its characteristics. However, scalability for reactive moving actor applications does not come by default. Specifically, to provide scalability and reliability, physical instantiations of actors in Orleans are distributed across the silo instances~\citep{bernstein2014orleans}. Actors are activated on silos based on the calculation of an actor placement strategy. However, as \citet{DBLP:conf/eurosys/NewellKMGAS16} pointed out, Orleans built-in actor placement policies are insufficient to achieve scalability, which can also be seen in our experiments, particularly in Figure~\ref{fig:scaleoutT}. For instance, the default RandomPlacement strategy results in good load balancing, but this strategy ignores spatial locality in interactions among moving actors. The latter leads to unnecessary latency and overheads, which significantly degrades system scalability. These effects are avoided in Dolphin by our spatially aware system-level actor design, which leverages spatial partitioning and indexing techniques not available in Orleans. Moreover, existing actor programming models, including Orleans, lack built-in support -- with well-defined concurrency semantics -- for the reactive data management functionalities required by reactive moving object applications, complicating the development of these applications.

Previous investigations about location-aware sensors, devices, and infrastructures have studied how to represent, manage, and query moving objects (\textit{e.g.},~\citet{GutingSchneider2005,DBLP:journals/vldb/GutingAD06}), as well as how to define and implement correct behaviors of moving object databases (\textit{e.g.}, \citet{DBLP:conf/icde/LuG14, DBLP:conf/ssdbm/WolfsonXCJ98}). Such studies on moving object databases are mainly focused on data structures, algorithms, and architectures for supporting past, present, and near-future queries \citep{DBLP:journals/tods/PelanisSJ06,DBLP:journals/vldb/SidlauskasSJ14,DBLP:journals/pvldb/SowellSCDG13,DBLP:conf/vldb/JensenLO04,DBLP:conf/vldb/JensenP07}. 
However, reactive moving object applications require novel reactive API as demonstrated by our Moving Actor abstraction, which is not supported in existing moving object databases and libraries. 

RxSpatial~\citep{DBLP:conf/icde/HendawiGSFA17} is probably one of the only works toward reactive spatial data management. RxSpatial married the Microsoft SQL Server Spatial Library \citep{msspatial} with ReactiveX~\citep{msreactivex} to implement a real-time reactive spatial library. RxSpatial provides a reactive spatial API, employs a spatial index structure for moving object queries, and continuously monitors and detects the intersection and distance between moving objects~\citep{DBLP:conf/sigmod/ShiHFA16, DBLP:conf/gis/ShiHGFA16, DBLP:conf/debs/HendawiSFKA16}. However, the reactivity model in RxSpatial is limiting in terms of supporting reactive moving object applications. Reactive moving objects in RxSpatial need to explicitly subscribe to other objects in order to monitor the relation between them, which is not an efficient solution. For example, in order to implement the the reactive sensing API in our platform, every reactive moving object would have to subscribe to the movements of all the other moving objects in the system, which could be highly expensive. Dolphin, by contrast, allows reactive moving objects to subscribe to intensional representations in terms of fences and spatial predicates, obviating explicit object-to-object subscriptions and enabling a partitioned actor-based design.


There are also some solutions that provide spatial event monitoring by using stream processing systems. The studies conducted in~\citep{DBLP:conf/gis/AliCRK10, DBLP:conf/ssd/MillerRAAHKLZTA11, DBLP:journals/pvldb/KazemitabarDAAS10, DBLP:journals/debu/AliCRK10} support continuous spatial queries for monitoring. \citet{DBLP:journals/geoinformatica/GalicMO17} proposed a distributed spatio-temporal mobility data stream framework to support continuous queries over streams. GeoFlink~\citep{DBLP:conf/cikm/ShaikhMKK20} extended Apache Flink to support spatial data types, indexes, and continuous queries over spatial data streams. The stream processing abstraction in these systems, often in the form of a topology of stream operators, is suitable for specifying a set of bulk operations over a large number of data items, such as transformation, filtering, join, and aggregate. However, it is unnatural and difficult to use this abstraction to specify complex application logic, such as heterogeneous behaviors of reactive moving objects. The latter creates an impedance for practitioners to implement the application tier of reactive applications. Additionally, stream processing systems do not naturally fulfill the design objective O1 in Section~\ref{sec:reqs}, which introduces complexities for developers in plugging together the concurrency semantics of these engines with those of spatially aware data stores. By contrast, Dolphin enriches the popular actor programming model with features of reactive moving object data management as well as with principled actor-based spatiotemporal concurrency semantics covering movement, spatial queries, and reactions. Furthermore, Dolphin's fully spatially partitioned design allows us to combine virtual actors with the Orleans streams APIs into a novel and scalable implementation supporting all of the aforementioned features. 


\section{Conclusions} 
\label{sec:conclusion}
Emerging moving object applications, such as Cooperative Intelligent Transportation Systems, require additional support for reactivity based on spatial sensing of dynamic data. Furthermore, these applications require scalability, elasticity and low-latency, all of which affect programmer productivity. 
We proposed a scalable, elastic, reactive, spatial data caching layer for moving object applications to meet these requirements holistically. 
We posited a new data management system architecture for reactive moving object applications by adding reactive moving object data management functionality in actor-oriented databases. 
We judiciously designed a new programming model for this new class of moving actor-oriented databases and carefully implemented it in a system called Dolphin, which extends the Microsoft Orleans virtual actor runtime. 
Our experiments using both micro-benchmarks and a C-ITS scenario benchmark showed that Dolphin meets the real-time performance requirements for spatial data ingestion, querying, and reactivity. 
Moreover, our experiments highlighted that our system seamlessly 
scales out across multiple machines. 

We hope that Dolphin, for which we plan to release artifacts upon acceptance, can help alleviate the complexity of development and data management for reactive moving object applications. Additionally, given the huge potential for societal impact of these mobile Internet-of-Things applications, we aspire that our work may serve as a catalyst to further motivate the research community to explore the possibilities and challenges of enriching actor runtimes with programming abstractions and algorithms tailored to this scenario. 



\bibliographystyle{ACM-Reference-Format}
\bibliography{bib-sigmod}

\end{document}